\documentclass{elsarticle}

\usepackage{lineno,hyperref}
\usepackage{graphicx}
\usepackage{epstopdf}
\usepackage{float}
\usepackage{natbib}
\usepackage{amsthm}
\usepackage{amssymb,bm}
\usepackage{latexsym}
\usepackage[title]{appendix}
\usepackage{makecell}
\usepackage{lscape}
\usepackage{pdflscape}

\usepackage{mathptmx}
\usepackage{setspace}
\usepackage{color}

\usepackage{algorithm}
\usepackage{algpseudocode}

\usepackage[fleqn]{amsmath}
\usepackage{amsfonts}
\usepackage{amsmath}
\usepackage{multirow}
\usepackage{amsthm}

\biboptions{authoryear}

\begin{document}

\begin{frontmatter}

\title{Learning-based model predictive control for passenger-oriented train rescheduling with flexible train composition}


\author[mymainaddress]{Xiaoyu Liu\fnref{fn1}}
\ead{X.Liu-20@tudelft.nl}

\author[mymainaddress]{Caio~Fabio~Oliveira~da~Silva\fnref{fn1}}
\ead{c.f.oliveiradasilva@tudelft.nl}

\author[mymainaddress]{Azita Dabiri}
\ead{a.dabiri@tudelft.nl}

\author[mysecondaryaddress]{Yihui Wang}
\ead{yihui.wang@bjtu.edu.cn}

\author[mymainaddress]{Bart De Schutter}
\ead{b.deschutter@tudelft.nl}

 \fntext[fn1]{X Liu and CFO~da~Silva contributed equally.}

\address[mymainaddress]{Delft Center for Systems and Control, Delft University of Technology, 2628 CD Delft, The Netherlands}
\address[mysecondaryaddress]{School of Automation and Intelligence, Beijing Jiaotong University, Beijing 100044, China}

\begin{abstract}
This paper focuses on passenger-oriented real-time train rescheduling, considering flexible train composition and rolling stock circulation, by integrating learning-based and optimization-based approaches. A learning-based model predictive control (MPC) approach is developed for real-time train rescheduling with flexible train composition and rolling stock circulation to address time-varying passenger demands. In the proposed approach, the values of the integer variables are obtained by pre-trained long short-term memory (LSTM) networks, while the continuous variables are determined through nonlinear constrained optimization.  The learning-based MPC approach enables us to jointly consider efficiency and constraint satisfaction by combining learning-based and optimization-based approaches. In order to reduce the number of integer variables, four presolve techniques are developed to prune a subset of integer decision variables. 
Numerical simulations based on real-life data from the Beijing urban rail transit system are conducted to illustrate the effectiveness of the developed learning-based MPC approach.
\end{abstract}

\end{frontmatter}


\section{Introduction}
Urban rail transit has become increasingly important in large cities due to its reliability, high capacity, and eco-friendly characteristics. Urban rail transit systems prioritize safe and efficient train operations while providing high-quality service to passengers.  
Effective real-time train scheduling is essential for enhancing passenger satisfaction and minimizing operational costs within infrastructure limitations. However, the rapid expansion of urban rail transit systems and the increasing passenger demands pose significant challenges to real-time scheduling. Advanced scheduling models and control strategies are required to develop efficient timetables and to improve the overall performance of urban rail transit systems. 

\subsection{Passenger-oriented train scheduling}\label{passenger-oriented}
In urban rail transit systems, passenger demands vary throughout the day, necessitating adjustments of the train schedules to accommodate these demand variations while considering operational costs. 
One direction addresses time-varying passenger demands by optimizing the departure and arrival times of trains at each station, while taking into account several attributes of train operations and infrastructure restrictions, e.g., train stopping plans (the set of station stops for each train) \citep{cacchiani2020robust,qi2021integer}, rolling stock circulations \citep{wang2018passenger}, and train speed levels \citep{hou2019energy,wang2021energy}.   \cite{qi2021integer} optimized train stopping plans and timetables of a high-speed railway line considering time-varying passenger demands by formulating a mixed-integer linear programming (MILP) problem, and the aim is to find a solution that considers passenger preferences for departure times while ensuring trains operate within capacity limits. 
Considering the passenger load of trains, \cite{wu2021multi} minimized the passenger waiting time and the energy consumption by developing a heuristic algorithm to solve the resulting nonlinear integer programming problem. \cite{wang2021energy} formulated a mixed-integer nonlinear programming (MINLP) problem to minimize train energy consumption and passenger waiting times by optimizing train speed levels and headway deviations. To address train scheduling under uncertain passenger demands,  \cite{cacchiani2020robust} introduced a protection level on the departure and arrival times, and solved an MILP to obtain the train schedule and the stopping plan. \cite{huang2025integrated} investigated jointly minimization of operation costs, passenger waiting time, and passenger travel time, and developed a variable neighborhood search approach to solve the resulting problem. However, the above studies optimize train schedules within the fixed transport capacity of a line, such as fixed train compositions or fixed train departure frequencies. As transport capacity directly impacts passenger flows, further research is required to improve passenger satisfaction by including transport capacity as a decision variable.

Another direction for train scheduling problems addresses time-varying passenger demands by optimizing transport capacity explicitly.  Several studies focus on optimizing transport capacity by adjusting train departure frequencies, with higher frequencies during peak hours and lower frequencies during off-peak hours \citep{canca2016setting,liu2023bi,liu2024real}. To handle uncertain passenger demands, \cite{liu2024real} developed a scenario-based distributed control approach, in which a scenario reduction technique is applied to reduce the computational burden. \cite{pu2021two} developed a two-step approach in which the first step determines the nominal line plan and the second step adjusts the time plan based on the actual measured passenger demands.  
Instead of changing the departure frequency, which would significantly affect the schedule, in recent years, many researchers have focused on optimizing the composition of the train \citep{ying2022adaptive,zhao2023integrated}. \cite{pan2023demand} developed a column-generation-based approach to optimize the timetable, train composition, and rolling stock circulation plan of an urban rail transit line. Their paper concludes that flexible train composition can provide additional adaptability to better match time-varying passenger demands. \cite{wang2024flexible} investigated flexible train composition and rolling stock circulation of an urban rail transit line, and solved the resulting MILP problem by developing an approximation approach and a two-stage meta-heuristic algorithm. \cite{yang2024integrated} investigated the train scheduling problem with flexible train composition for an urban rail transit line, and they applied an adaptive large neighborhood search algorithm to solve the resulting integer programming problem. As the inclusion of train composition optimization and rolling stock circulation planning introduces additional integer variables, the above studies indicate that the online computational complexity is the main challenge of incorporating flexible train composition into the real-time train scheduling problem.

\subsection{MPC for real-time train scheduling}\label{MPC-scheduling}
Model predictive control (MPC) has been widely adopted in various applications for its ability to handle multivariable constrained control problems \citep{grune2017nonlinear,mayne2000constrained,qin2003survey}. 
The train scheduling problem is a typical constrained control problem, and many studies have applied MPC for real-time train scheduling. \cite{de2002model} first applied MPC in the train scheduling problem to minimize train delays by adjusting transfer connections. \cite{caimi2012model} developed an MPC algorithm to optimize timetables, transfer connections, and train assignment plans in complex station areas.  \cite{cavone2020mpc} proposed an MPC approach for train rescheduling during disruptions and delays, where in each step the resulting MILP problem is solved by combining bi-level heuristics and distributed optimization. The above studies only handle operator-related factors, leaving an open gap in including passenger demands in real-time train scheduling problems to improve the service quality of urban rail transit systems.

In recent years, several studies have focused on MPC for real-time passenger-oriented train scheduling. \cite{assis2004generation} applied MPC to compute the train timetable of a metro line considering train headway and passenger load, where a linear programming problem is solved at each step.  Based on a state-space model, \cite{li2016robust} developed a robust MPC approach to minimize the upper bound of the timetable deviation from the nominal timetable under uncertain disturbances. By solving linear matrix inequalities, they constructed a Lyapunov function to ensure the attenuation of the timetable deviation. An event-triggered MPC approach is further developed by \cite{wang2022event} to reduce the computational burden of updating control variables in each step. However, these studies do not explicitly consider train capacity limitations and time-varying passenger demands, and the results are based on the assumption that the maximum number of passengers does not exceed the maximum train capacity. \cite{chen2022real} investigated real-time train rescheduling and stop-skipping strategies under disturbances and dynamic passenger flows. A mixed-integer quadratic programming (MIQP) problem is solved at each MPC step to minimize the total deviation from the original timetable and the number of waiting passengers. In the follow-up work, \cite{chen2023distributed} applied distributed MPC to address the real-time train regulation problem, using Dantzig-Wolfe decomposition to solve the resulting optimization problem.  
\cite{liu2023modeling} explicitly incorporated time-varying passenger demands and train capacity limitations into the real-time train scheduling problem. In \cite{liu2023modeling}, the time-varying passenger demand is approximated as a piecewise constant function by dividing the planning time window into several time intervals of equal length, and an MILP-based MPC approach is adopted for real-time train scheduling. 
The main challenge of applying MPC in real time is the online computational burden. Including additional attributes, such as train capacity, train composition, and rolling stock circulation, will further increase the computational burden. Therefore, further research is required to develop efficient MPC approaches for real-time passenger-oriented train scheduling. 

\subsection{Learning-based train scheduling}\label{learning-scheduling}
Learning-based approaches are considered effective methodologies for reducing the online computational burden. Learning-based approaches, including deep learning and reinforcement learning (RL), have also been applied in train scheduling problems in recent years \citep{tang2022literature}. 
SL trains models on labeled data to make accurate predictions or classifications, while RL trains an agent to make decisions through trial-and-error using rewards and penalties. \cite{kuppusamy2020deep} applied a deep learning approach to an energy-efficient timetable rescheduling problem, where a long short-term memory (LSTM) network is trained to select the optimal operation mode. 
\cite{vsemrov2016reinforcement} applied Q-learning in the train rescheduling problem to reduce delays caused by disturbances and disruptions, and they illustrated their method on a single-lane track with three trains. 
\cite{yin2016energy} proposed an approximate dynamic programming approach to address train rescheduling problems, aiming to reduce passenger delays, total travel time, and train energy consumption. In \cite{yin2016energy}, the states include disturbance information, train arrival times, the number of boarding passengers, and the number of waiting passengers, while the actions include rescheduling the dwell times and running times. \cite{khadilkar2019scalable} applied RL to determine track allocations and timetables of bidirectional railway lines to minimize the priority-weighted delay. 
\cite{ying2022adaptive} developed a proximal policy optimization approach for the train scheduling problem in an urban rail transit line, considering flexible train composition. In \cite{ying2022adaptive}, the control policy and the value function are parameterized by artificial neural networks, and scheduling constraints are handled by a devised mask scheme. Simulation results show that this approach reduces the computational burden and improves solution quality compared to the genetic algorithm and differential evolution.
More studies of learning-based approaches in railway systems can be found in the recent review paper \citep{tang2022literature}.

In the above studies, scalability and constraint satisfaction are two main challenges in developing learning-based train scheduling approaches. 
The train scheduling problem is typically formulated as an MILP or MINLP problem, and the computational complexity increases rapidly as the number of integer variables increases. In recent years, some research has combined learning-based and optimization-based approaches for MILP or MINLP problems by using learning-based approaches to obtain the integer variables. Having the integer variables fixed, the continuous variables are then obtained by solving a linear or nonlinear programming problem. In this context, the aim is to combine the advantages of both learning-based and optimization-based approaches, i.e., the online computational efficiency of learning-based approaches and the constraint satisfaction of optimization-based approaches \citep{cauligi2022prism,russo2023learning}. Promising results of learning-based approaches in railway systems and the novel learning-based approaches in mixed integer programming problems have inspired us to develop new learning-based frameworks for real-time train scheduling.

\begin{landscape}
\begin{table}[h]
		\centering
		\begin{tabular}{llllllll}\hline
  Publications&Infrastructure&\makecell[l]{Train\\ composition }&\makecell[l]{Train \\order}&\makecell[l]{\small{Rolling stock}\\ \small{circulation}}&\makecell[l]{Passenger\\ demands}&\makecell[l]{Solution method}& Objective (s)\\\hline
  \cite{vsemrov2016reinforcement}& single line& fixed & yes & no & no & Q learning&  minimize the total delay of trains\\%
  \cite{yin2016energy}& single line& fixed & no & no & yes& \makecell[l]{approximate dynamic programming}& \makecell[l]{minimize delays of affected passengers,\\ total travel time of passengers,\\ and the train energy consumption}\\
  \cite{li2016robust}& single line &   fixed & no & no & yes & \makecell[l]{robust MPC} & \makecell[l]{minimize the deviations \\form the nominal timetable}\\ 
  \cite{khadilkar2019scalable} & single line& fixed& yes & no & no & Q learning  & minimize the priority-weighted delay\\ 
  \cite{cacchiani2020robust} & single line & fixed & yes & no & yes & robust MPC & \makecell[l]{minimize train travel time and \\unsatisfied passenger demands} \\
  \small{\cite{kuppusamy2020deep}}&single line&  fixed & no & no & no & deep learning & minimize energy consumption\\ 
  \cite{pu2021two} & single line& fixed & no & no & yes &Lagrangian relaxation& \makecell[l]{minimize operational costs and \\total passenger travel time} \\ 
  \cite{wang2021energy} & single line& fixed & no & yes & yes & MILP & \makecell[l]{minimize train energy consumption \\ and passenger waiting times}\\ 
  \cite{ying2022adaptive} & single line& flexible& no & yes & yes & deep reinforcement learning& \makecell[l]{minimize the operation cost and\\ passenger waiting times}\\
  \cite{cavone2020mpc} & network & fixed & yes & no & no & bi-level heuristics, MPC &  minimize train delays \\
  \cite{wang2022event} & single line& fixed & no & no& yes  & event-triggered MPC& \makecell[l]{minimize the deviations \\form the nominal timetable}\\ 
  \cite{chen2022real} & single line& fixed & no & yes & yes & MPC & \makecell[l]{ minimize total train deviation and\\  the number of waiting passengers}\\ 
  \cite{liu2023modeling}& network& fixed & no & no& yes &  MPC& minimize the  waiting time of passengers\\ 
  \cite{pan2023demand}& single line& flexible& yes& yes& yes & column-generation-based heuristic& \makecell[l]{minimize passenger waiting times and\\ the usages of train units}\\ 
  \cite{liu2023bi}& network& fixed & yes& yes& yes& bi-level MPC& \makecell[l]{minimize passenger travel times and \\ the total operation costs}\\ 
  \cite{liu2024real}& network & fixed & no & yes & yes & scenario-based  distributed MPC& \makecell[l]{minimize passenger travel times and \\ the total operation costs}\\ 
  \cite{wang2024flexible}& single line& flexible & no & yes & yes & two-stage heuristic & \makecell[l]{ reduce the number of stranded passengers, \\ the number of rolling stocks, \\ and the operational costs}\\ 
  \cite{yang2024integrated} & single line& flexible & no & yes & yes & adaptive large neighborhood search & \makecell[l]{minimize total
waiting time of passengers\\ and the number of train units used}\\ 
\cite{huang2025integrated}& single line& fixed & no& yes & yes & variable neighborhood search & \makecell[l]{minimize passenger waiting time, \\ travel time and operation costs}\\
      Current paper& network & flexible & yes& yes & yes & deep-learning-based MPC & \makecell[l]{minimize passenger waiting times \\ and operation costs} \\ 
\hline
		\end{tabular}
		\caption{Summary of relevant studies on passenger-oriented train rescheduling.}
		\label{literature}
	\end{table}
\end{landscape}


\subsection{Paper contributions and structure}

The studies introduced in Sections~\ref{passenger-oriented}, \ref{MPC-scheduling}, and \ref{learning-scheduling} are summarized in Table~\ref{literature}, categorized by network details, train composition, train order, rolling stock circulation, passenger demands, and solution method. 
From the above literature review, it can be found that flexible train composition combined with rolling stock circulation planning is an effective approach to balance operational costs and service quality. Shorter trains can be deployed during off-peak hours and longer trains during peak hours, thereby balancing operational costs and passenger waiting times. The passenger-oriented train scheduling problem is typically formulated as a mixed-integer programming problem. Various approaches can be applied to solve the resulting scheduling problem. Among them, optimization-based methods often encounter computational challenges for large-scale networks, necessitating additional efforts such as heuristics, two-step frameworks, and distributed approaches to reduce the computational burden. Learning-based methods are considered promising for addressing computational complexity. However, available learning-based methods often face scalability challenges, and existing research is generally limited to small-scale scenarios involving a single rail line. 

The current paper, therefore, addresses the learning-based train rescheduling problem in rail networks, taking into account time-varying passenger demands, flexible train composition, and rolling stock circulation. Unlike existing train rescheduling methods in the literature, the proposed approach integrates optimization and deep learning to combine the constraint satisfaction advantages of optimization-based methods with the fast online computation advantages of learning-based approaches. 
The main contributions of the paper are listed as follows:
\begin{enumerate}
    \item A passenger-oriented train rescheduling model in \cite{liu2023modeling} is extended to include time-varying sectional passenger demands, flexible train composition, and rolling stock circulation. The time-varying passenger demands can be approximated as a piecewise constant function by dividing the prediction horizon into several time intervals.
    \item Four presolve techniques are developed to streamline optimization processes by pruning a subset of integer decision variables. After implementing the presolve techniques,  long short-term memory (LSTM) networks are applied to obtain the remaining integer variables with reduced dimensions. 
    \item A deep-learning-based MPC approach is developed for real-time train rescheduling. To improve the online computational efficiency of MPC, the learning-based approach is applied to obtain integer variables while the detailed timetable is obtained by solving a constrained optimization problem with the fixed integer variables. 
\end{enumerate}

The remaining part of the paper is organized as follows: Section~\ref{lmpc-section2} provides the problem description and general explanations. Section~\ref{lmpc-section3} introduces the passenger-oriented train rescheduling model. In Section~\ref{lmpc-mpc_approach}, the problem formulation and the MINLP-based MPC approach are presented. In Section~\ref{lmpc-lmpc_approach}, we propose a learning-based MPC approach for real-time train rescheduling. Section~\ref{lmpc-section6} provides an illustrative case study. Section~\ref{lmpc-section7} concludes the paper.

\section{Problem description and explanations}\label{lmpc-section2}

\begin{figure}[htbp]
\begin{center}
\includegraphics[width=9.5cm]{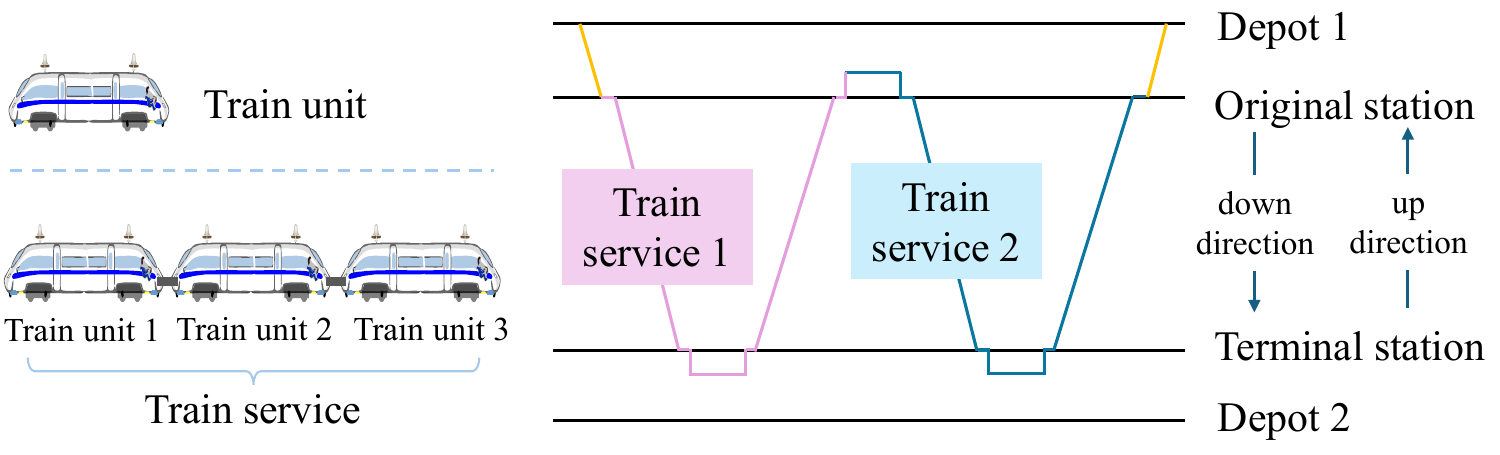}    
\caption{Illustration of definitions used in this paper.}
\label{lmpc-train_composition}
\end{center}
\end{figure}

In urban rail transit systems, a line is defined as the route of trains with the same origin, intermediate, and destination stations. A train service is defined as a train departure from its origin station, visiting every station in the line, and finally returning to the depot (or connecting to other train services at the depot). As illustrated in Fig.~\ref{lmpc-train_composition}, a train service consists of one or several train units, and the composition can be changed at the station connected to a depot by adding or removing train units. 

In this paper, we jointly optimize train compositions and timetables considering time-varying passenger demands and rolling stock circulation. We focus on the train rescheduling problem with the objective of optimizing departure times, arrival times, and train composition. In this context, both train travel times and transport capacity are adjusted to accommodate the time-varying passenger demands.  We aim to minimize the total waiting time of passengers and the train energy consumption, and the control actions are the train composition, train service orders at the platforms connected to the same depot, and departure/arrival times.  Flexible train composition and rolling stock circulation relate to the set of integer variables, while train timetables relate to the set of continuous variables, such as departure and arrival times. Trains operate under several constraints, including train capacity constraints, train availability in a depot, and headway constraints. By applying MPC, we solve the resulting mixed-integer programming problem in a moving horizon manner for real-time train rescheduling. To improve the online computational efficiency of MPC, we use the learning-based approach to obtain integer variables, i.e., train compositions and train orders; then, we optimize the detailed timetable with the fixed integer variables by solving a constrained optimization problem.

\section{Mathematical formulation for passenger-oriented train rescheduling}\label{lmpc-section3}
In this section, we develop a passenger-oriented train rescheduling model for urban rail transit systems.  The notations for the model formulation are provided in Section~\ref{lmpc-section3.1}.  The train operation constraints of the model are introduced in Section~\ref{lmpc-section3.2}. In Section~\ref{lmpc-section_roll}, the rolling stock circulation constraints related to the model are introduced. In Section~\ref{lmpc-section_passenger}, passenger flow constraints of the model are presented.



\subsection{Notations}\label{lmpc-section3.1}

Tables \ref{input parameters}, \ref{decision variables}, and \ref{output parameters} list the indices and input parameters, the decision variables, and the output variables of the model formulations, respectively.

 \begin{table}[h]
		\centering
		\begin{tabular}{ll}\hline
		Notations&Definition\\\hline
		$p$&Index of platforms, $p \in \cal{P}$; $\cal{P}$ is the set of platforms\\
		$k_p$&Index of train services at platform $p$, $k_p \in \mathcal{I}_p$; $\mathcal{I}_p$ is the set of train services departing from platform $p$ \\
        $z$& Index of depots \\
		$\mathrm{p}^\mathrm{pla}\left( p \right)$& Predecessor platform of platform $p$ \\
        $\mathrm{s}^\mathrm{pla}\left( p \right)$& Successor platform of platform $p$\\
        $d^\mathrm{pre}_{p}(k_p)$ & Predetermined departure time of train service $k_p$ at platform $p$\\
        $h_{p}^\mathrm{min}$ & Minimum departure-arrival headway at platform $p$\\
        $r_p^\mathrm{min}$  & Minimum running time of trains from platform $p$ to its succeeding platform\\
        $ r_p^\mathrm{max}$ & Maximum running time of trains from platform $p$ to its succeeding platform\\
        $C_\mathrm{max}$ & Maximum capacity of a train unit\\
        $\ell^\mathrm{min}$ & Minimum number of train units allowed to be included in any train service \\
		$\ell^\mathrm{max}$ & Maximum number of train units allowed to be included in any train service\\
        $\sigma_p$ & Parameter indicating whether the train composition can be adjusted at platform $p$,  \\
        & i.e., if train composition can be adjusted at platform $p$, then, $\sigma_p = 1$, otherwise, $\sigma_p = 0$.\\
        $\tau_{p}^\mathrm{min}$ & Minimum dwell time of a train service at platform $p$\\
        $t_{p}^\mathrm{cons}$ & Time required for changing the train composition at platform $p$\\
        $t_{p}^\mathrm{roll}$ & Time for trains from platform  $p$ to other platforms corresponding to the same depot\\
        $\mathrm{dep}(z)$ & Set of platforms directly connected with depot $z$\\
        $\mathrm{pla}(p)$ & Set of platforms belonging to the same station as platform $p$\\
        ${\rho_{p} (k_p)}$ & Passenger demands from platform $p$ to its successor platform during $d^\mathrm{pre}_{k-1,p}$ to  $d^\mathrm{pre}_{k,p}$\\
        $N_z^{\mathrm{train}}$& Total number of train units available at depot $z$\\
        $\chi_{k_q,q,k_p,p}$ & Binary parameter, which can be determined based on the predetermined timetable; if train $k_q$ \\
        & at platform $q$ has transfer connection with train $k_p$ at platform $p$, $\chi_{k_q,q,k_p,p} = 1$; otherwise, $\chi_{k_q,q,k_p,p} = 0$.\\
        $\beta_{q,p}$ & Transfer rate from platform $q$ to platform $p$\\
        $t_{q}^\mathrm{trans}$ & Average transfer time from platform $q$ to the platforms at the same station\\
        $E_p^\mathrm{energy}$ & Average energy consumption for a train unit running from platform $p$ to its successor platform\\
        ${E_p^\mathrm{add}}$ & Additional cost of changing train composition at platform $p$\\

\hline
		\end{tabular}
		\caption{Indices and input parameters.}
		\label{input parameters}
	\end{table}

\begin{table}[h]
\centering
\begin{tabular}{ll}\hline
Notations&Definition\\\hline
$d_{p}(k_p)$ & Departure time of train service $k_p$ at platform $p$\\
$a_{p}(k_p)$ & Arrival time of train service $k_p$ at platform $p$\\
$\ell_{p}(k_p)$ & Number of train units included in train service $k_p$ at platform $p$, $\ell_{p}(k_p) \in \mathbb{Z}_+$\\
${\xi_{k_p,k_{p'},p,p'}}$ & Binary variable; if the train units from train service $k_{p'}$ at platform $p'$ can be used for train service $k_{p}$ \\
 & at platform $p$,  ${\xi_{k_p,k_{p'},p,p'}} = 1$; otherwise, ${\xi_{k_p,k_{p'},p,p'}} = 0$.\\
$y_{p}(k_p)$ & Number of train units coming to/from the depot for train service $k_p$, $y_{p}(k_p) \in \mathbb{Z}$\\
$\eta_{k_p,p}$ & Binary variable; if the composition of train service $k_p$ is changed at platform $p$, $\eta_{k_p,p} = 1$;\\
 & otherwise, $\eta_{k_p,p} = 0$.\\
\hline
\end{tabular}
\caption{Decision variables}
\label{decision variables}
\end{table}

\begin{table}[h]
\centering
\begin{tabular}{ll}\hline
Notations&Definition\\\hline
$\tau_{p}(k_p)$ & Dwell time of train service $k_p$ at platform $p$ \\
$h_{p}(k_p)$ & Departure-arrival headway between train service $k_p$ and train service $k_p+1$ at platform $p$\\
$r_{p}(k_p)$ & Running time of train service $i$ from platform $p$ to its successor platform \\
$r_{p}^\mathrm{turn}(k_{p})$ & Turnaround time of train service $k$ at platform $p$\\
$\tau^\mathrm{add}_{p}(k_p)$ & Additional time required for changing the composition of train service $k_p$ at platform $p$\\
$n_{p}(k_p)$ & Number of passengers waiting at platform $p$ at time $d^\mathrm{pre}_{p}(k_p)$\\
$ n^\mathrm{trans}_{p}(k_p)$ & Number of transfer passengers arriving at platform $p$ for train $k_p$\\
$n^\mathrm{depart}_{p}(k_p)$ & Number of passengers departing from platform $p$ with train service $k_p$\\
$n^\mathrm{arrive}_{p}(k_p)$ & Number of passengers arriving at platform $p$ with train $k_p$ from the predecessor platform $\mathrm{p}^\mathrm{pla}\left( p \right)$\\
$n^\mathrm{before}_{p}(k_p)$ & Number of passengers waiting at platform $p$ immediately before the departure of train service $k_p$\\
$C_{p}(k_p)$ & Total capacity of train service $k_p$ at platform $p$\\
$n^\mathrm{after}_{p}(k_p)$ & Number of passengers waiting at platform $p$ immediately after train service $k_p$ departs from platform $p$.\\
\hline
\end{tabular}
\caption{Output variables}
\label{output parameters}
\end{table}

\subsection{Train operation constraints}\label{lmpc-section3.2}
\begin{figure}[htbp]
\begin{center}
\includegraphics[width=9.5cm]{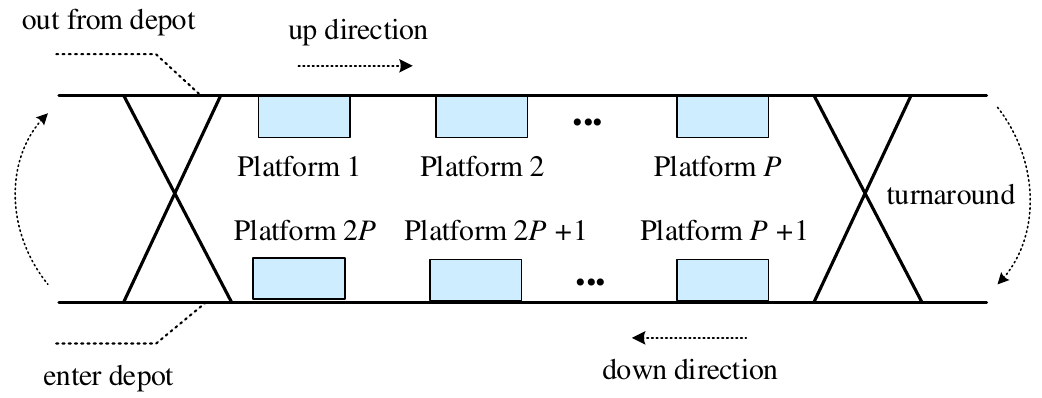}    
\caption{Layout of a bidirectional urban rail transit line.}
\label{lmpc-line}
\end{center}
\end{figure}

In urban rail transit systems, each line typically comprises two directions, i.e., the up direction and the down direction, as shown for a line of $P$ stations in Fig.~\ref{lmpc-line}. For each line, a train service can be defined as a train running from the starting platform to the terminal platform, e.g., from Platform 1 to Platform 2$P$ in Fig.~\ref{lmpc-line}.
Trains generally operate following a predetermined timetable, and the predetermined departure time of train service $k_p$ at platform $p$ is represented by $d^\mathrm{pre}_{p}(k_p)$. 

In practice, premature departure is usually not permitted; thus, the predetermined departure time defines a lower bound of the actual departure time. In general, passengers expect regular departures at every platform so that they can conveniently plan their travels and avoid extended waiting times for the next train in case they miss a connecting train. Therefore, in this paper, we do not change the departure frequency of trains when adjusting the timetable and train composition. Hence, the actual departure time is constrained by
\begin{equation} \label{lmpc-predetermined}
d^\mathrm{pre}_{p}(k_p) \le d_{p}(k_p) < d^\mathrm{pre}_{p}(k_p+1),
\end{equation}
where $d_{p}(k_p)$ is the actual departure time of train service $k_p$ at platform $p$ determined by 
\begin{align}
 &d_{p}(k_p) = a_{p}(k_p) + \tau_{p}(k_p),
\end{align}
where $a_{p}(k_p)$ and $\tau_{p}(k_p)$ are the arrival time and dwell time of train service $k_p$ at platform $p$. 

For the safe operation of trains, the headway constraint should be satisfied:
\begin{align}
 &a_{p}(k_p+1) = d_{p}(k_p) + h_{p}(k_p),\\
 &h_{p}(k_p) \ge h^\mathrm{min}_{p},
\end{align}
where $h_{p}(k_p)$ is headway of train service $k_p$ at platform $p$ , and $h^\mathrm{min}_p$ denotes the minimum headway.

The arrival time of train service $k_p$ at the successor platform of platform $p$ should also satisfy
\begin{align}
 &a_{\mathrm{s}^\mathrm{pla}\left( p \right)}(k_p) = d_{p}(k_p) + r_{p}(k_p),\\
 &r_{p}^\mathrm{min} \le r_{p}(k_p) \le r_{p}^\mathrm{max},
\end{align}
where $r_{p}(k_p)$ is the running time of train service $k_p$ from platform $p$ to platform $\mathrm{s}^\mathrm{pla}\left( p \right)$, and $r_{p}^\mathrm{min}$ and $r_{p}^\mathrm{max}$ are the minimum and maximum running time from platform $p$ to platform $\mathrm{s}^\mathrm{pla}\left( p \right)$.


\subsection{Rolling stock circulation constraints}\label{lmpc-section_roll}
At the terminal station, a turnaround action is required for the continuation of the train service. 
The turnaround constraints can be formulated as:
\begin{align}\label{lmpc-turnaround}
 &a_{\mathrm{s}^\mathrm{pla}\left( p \right)}(k_{\mathrm{s}^\mathrm{pla}\left( p \right)})= d_{p}(k_{p}) + r_{p}^\mathrm{turn}(k_{p}),\\
 &r_{p}^\mathrm{turn,\min} \le r_{p}^\mathrm{turn}(k_{p}) \le r_{p}^\mathrm{turn,\max},
\end{align}
where $r_{p}^\mathrm{turn}(k_{p})$ represents the turnaround time of train service $k$ at platform $p$, and $r_{p,\min}^\mathrm{turn}$ and $r_{p,\max}^\mathrm{turn}$ denote the minimum and maximum turnaround times at platform $p$. 

An urban rail transit line typically has a limited number of train units that either operate on the line or are stored in the depot. The train composition can be adjusted at the platform that is linked with the depot, and the number of train units $\ell_{p}(k_p) \in \mathbb{Z}_+$ for train service $k_p$ at platform $p$ is determined by
\begin{align}
&\ell_{p}(k_p) = \ell_{\mathrm{p}^\mathrm{pla}\left( p \right)}(k_{p}) + \sigma_p y_{p}(k_p), \label{lmpc-composition}\\
& \ell_{\min}\le \ell_{p}(k_{p}) \le \ell_{\max},
\end{align}
where $\sigma_p$ is a parameter related to the network layout indicating whether the train composition can be adjusted at platform $p$, i.e., if the train composition can be adjusted at platform $p$, then, $\sigma_p = 1$, otherwise, $\sigma_p = 0$. Moreover, $y_{p}(k_{p}) \in \mathbb{Z}$ represents the number of train units coming to/from the depot for train service $k$; specifically, if $y_{p}(k_{p}) > 0$, then, $y_{p}(k_{p})$ extra train units will come from the depot to be added to train service $k_p$; if $y_{p}(k_{p}) < 0$, train service $k_p$ will be decomposed and $|y_{p}(k_{p})|$ train units will return to the depot; if $y_{p}(k_{p}) = 0$ the composition of train service $k_p$ will not be changed at platform $p$. Furthermore, $\ell_{\min}$ and $\ell_{\max}$ represent the minimum and the maximum number of train units allowed to be included in any train service.


{\bf{Remark 1}}: 
If the depot is linked with the terminal platform (e.g., Platform 1 in Fig.~\ref{lmpc-line}) and train service $k$ is performed by the turnaround train units of the previous train service, then (\ref{lmpc-composition}) becomes $\ell_{1}(k_1) = \ell_{2P}(k_{2P}-1) +\sigma_1 y_{1}(k_1)$, i.e., we set $\ell_{1}(k_1) = \ell_{2P}(k_{2P}-1)$.

To capture the changes of train composition at platform $p$, we introduce a binary variable $\eta_{k_p,p}$ as
\begin{equation}\label{lmpc-eta_kp}
\eta_{k_p,p} = \begin{cases} 1, & \mathrm{if} \quad  |y_{p}(k_{p})| > 0;\\
0, & \mathrm{otherwise},
\end{cases}
\end{equation}
where $\eta_{k_p,p} = 1$ indicates the composition of train service $k_p$ is changed at platform $p$; otherwise, $\eta_{k_p,p} = 0$.

In general, additional dwell time is required when changing the train composition; thus, the dwell time $\tau_{p}(k_{p})$ of train service $k_p$ at platform $p$ should satisfy:
\begin{equation}\label{lmpc-tau_min}
\tau_{p}(k_{p}) \ge \tau^\mathrm{min}_{p} + \sigma_p \tau^\mathrm{add}_{p}(k_{p}),
\end{equation}
where $\tau^\mathrm{min}_p$ is the minimum dwell time at platform $p$, and $\tau^\mathrm{add}_{p}(k_{p})$ represents the additional time required for changing the composition of train service $k_p$ at platform $p$, which can be determined by 
\begin{equation}\label{lmpc-y_kp}
\tau^\mathrm{add}_{p}(k_{p}) = \eta_{k_p,p} \cdot t^\mathrm{cons}_{p},
\end{equation}
where $t^\mathrm{cons}_{p}$ is a constant representing the time required for changing the train composition at platform $p$.


A depot typically connects to at least one platform, and the departure order of train services at the corresponding platforms influences the number of available train units in a depot, i.e., if a newly arriving train service requires changing its composition, the total number of train units in the corresponding depot will change. To represent the relation of departure time of train services corresponding to the same depot, we define a binary variable ${\xi_{k_p,k_{p'},p,p'}}$ as 
\begin{equation}\label{lmpc-xi}
{\xi_{k_p,k_{p'},p,p'}} = \left\{ {\begin{array}{*{20}{l}}
{1,\quad {\rm{if}}\quad {d_{p}(k_{p})} \ge d_{p'}(k_{p'}) + t_{p'}^\mathrm{roll};}\\
{0,\quad {\rm{otherwise}},}
\end{array}} \right.
\end{equation}
where $d_{p}(k_{p})$ is the departure time of train service $k_p$ at platform $p$, ${d_{p'}(k_{p'})}$ is the departure time of train service $k_{p'}$ at platform $p'$, and $t_{p'}^\mathrm{roll}$ is the time for trains from platform $p'$ to other platforms corresponding to the same depot. In (\ref{lmpc-xi}), ${\xi_{k_p,k_{p'},p,p'}} = 1$ indicates train units in train service $k_{p'}$ from platform $p'$ is available for train service $k_p$ at platform $p$; otherwise, ${\xi_{k_p,k_{p'},p,p'}} = 0$.  

Then, the rolling stock circulation constraint is
\begin{equation}\label{lmpc-RS}
\small
 \sum_{k_p \in {\mathcal{I}}_{p}} {y_{p}(k_{p})} + \sum_{p' \in \mathrm{dep}(z)\backslash\{p\}} \sum_{k_{p'} \in {\mathcal{I}}_{p'}} \xi_{k_p,k_{p'},p,p'} {y_{p'}(k_{p'})} \le N_z^{\mathrm{train}},
\end{equation}
where $z$ is the index of the depot; $\mathrm{dep}(z)$ defines the set of platforms directly connected with depot $z$; $\mathcal{I}_{p'}$ defines the set of train services departing from platform $p'$; and $N_z^{\mathrm{train}}$ represents the total number of train units available at depot $z$. Eq.~(\ref{lmpc-RS}) indicates that the total number of train units departing from depot $z$ before time $d_{p}(k_{p})$ should be less than or equal to the total number of available train units in the depot. 

\subsection{Passenger flow constraints}\label{lmpc-section_passenger}
\begin{figure}[htbp]
\begin{center}
\includegraphics[width=9.5cm]{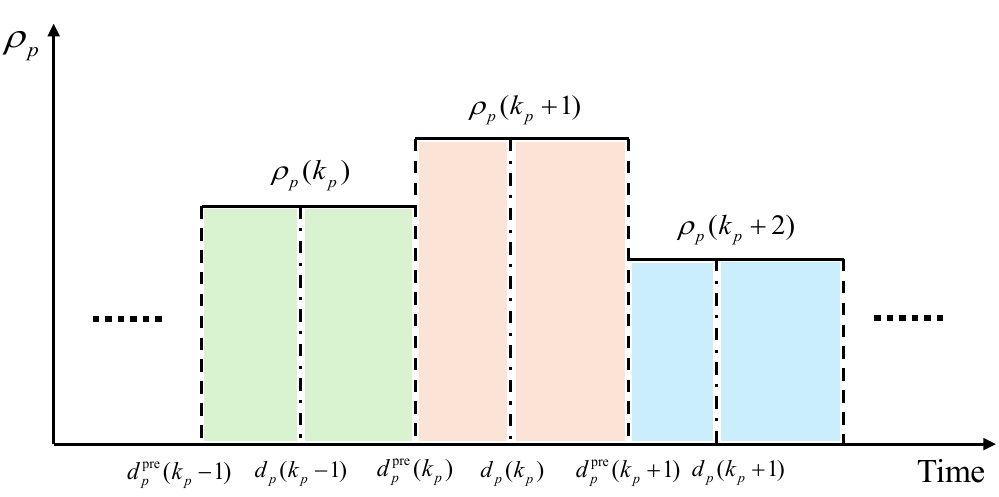}    
\caption{Illustration of piecewise approximation of passenger demands for platform $p$.}
\label{lmpc-demands_original}
\end{center}
\end{figure}

A predetermined timetable is generally designed based on the time-varying passenger demand to guide the daily operation of trains. At each platform, the predetermined timetable naturally divides the planning time window into several time intervals, 
with the predetermined train departure times at the platforms as the partition points. Following the previous research in \cite{liu2023modeling}, which demonstrated that the piecewise constant approximation of the time-varying passenger demand can significantly reduce computational complexity while preserving comparable accuracy, in this paper, we also approximate the passenger arrival rates as piecewise constant functions.  The resulting piecewise approximation is shown in Fig.~\ref{lmpc-demands_original} where $d^\mathrm{pre}_{p}(k_p)$ is the predetermined departure time of train service $k_p$; and $\rho_{p}(k_p)$ denotes passenger demands during $d^\mathrm{pre}_{p}(k_p-1)$ to  $d^\mathrm{pre}_{p}(k_p)$ for passengers aiming to leave platform $p$ with train service $k_p$. 

The number of passengers waiting at a platform immediately after the predetermined departure time of train service $k_p+1$ at platform $p$ can be calculated by
\begin{align}\label{lmpc-n}
n_{p}(k_p+1) = n_{p}(k_p) + \rho_{p}(k_p+1)\left(d^\mathrm{pre}_{p}(k_p+1) - d^\mathrm{pre}_{p}(k_p) \right) + n^\mathrm{trans}_{p}(k_p) - n^\mathrm{depart}_{p}(k_p),  
\end{align}
where $n_{p}(k_p)$ represents the number of passengers waiting at platform $p$ at time $d^\mathrm{pre}_{p}(k_p)$; $ \rho_{p}(k_p+1)\left(d^\mathrm{pre}_{p}(k_p+1) - d^\mathrm{pre}_{p}(k_p) \right)$ calculates the number of passengers arriving at platform $p$ between $d^\mathrm{pre}_{p}(k_p)$ and $d^\mathrm{pre}_{p}(k_p+1)$; $ n^\mathrm{trans}_{p}(k_p)$ denotes the number of transfer passengers arriving at platform $p$ for train $k_p$, and $n^\mathrm{depart}_{p}(k_p)$ denotes the number of passengers departing from platform $p$ with train service $k_p$. 



Since the departure time of each train service is adjusted according to (\ref{lmpc-predetermined}), there exists one departure in each time interval. As shown in  Fig.~\ref{lmpc-demands_original}, the number of passengers $n^\mathrm{before}_{p}(k_p)$ waiting at platform $p$ immediately before the departure of train service $k_p$ can be calculated by
\begin{align}\label{lmpc-w}
n^\mathrm{before}_{p}(k_p) = n_{p}(k_p) +  \rho_{p}(k_p+1) \left(d_{p}(k_p) - d^\mathrm{pre}_{p}(k_p) \right) + n^\mathrm{trans}_{p}(k_p) .
\end{align}


The number of passengers $n^\mathrm{depart}_{p}(k_p)$  departing from platform $p$ with train service $k_p$ should satisfy
\begin{subequations}\label{lmpc-boarding2}
\begin{align}
& n^\mathrm{depart}_{p}(k_p) \le C_{p}(k_p),\\
& n^\mathrm{depart}_{p}(k_p) \le n^\mathrm{before}_{p}(k_p) ,
\end{align}  
\end{subequations}
where $C_{p}(k_p)$ is the total capacity of train service $k_p$ at platform $p$. 


The total capacity of train service $k_p$ at platform $p$ is computed by 
\begin{align}
C_{p}(k_p) = \ell_{p}(k_p) C_\mathrm{max},
\end{align}
where $ \ell_{p}(k_p)$ is the number of train units composing train service $k_p$ at platform $p$.

The number of passengers $n^\mathrm{arrive}_{p}(k_p)$ arriving at platform $p$ with train $k_p$ from the predecessor platform $\mathrm{p}^\mathrm{pla}\left( p \right)$ can be calculated by
\begin{equation}
     n^\mathrm{arrive}_{p}(k_p) =  n^\mathrm{depart}_{\mathrm{p}^\mathrm{pla}\left( p \right)}(k_p).
\end{equation}
Then, the number of passengers $n^\mathrm{trans}_{p}(k_p)$ transferring to platform $p$ for train $k_p$ is computed by
\begin{equation}
     n^\mathrm{trans}_{p}(k_p) =  \sum_{q\in \mathrm{pla}(p)}{\sum_{k_q\in\mathcal{I}_q}{ \chi_{k_q,q,k_p,p} \beta_{q,p} n^\mathrm{arrive}_{q}(k_q)}},
\end{equation}
where $\mathrm{pla}(p)$ defines the set of platforms belonging to the same station as platform $p$, $\beta_{q,p}$ is the parameter defines the transfer rate from platform $q$ to platform $p$, and $\chi_{k_q,q,k_p,p}$ is the binary parameter denoting transfer connection between train $k_q$ at platform $q$ and train $k_p$ at platform $p$, which is defined as 
\begin{equation}\label{lmpc-chi}
\small
{\chi_{k_q,q,k_{p},p}} = 
\begin{cases}
    1, & \mathrm{if} \ {d_{p}^\mathrm{pre}(k_{p}-1)} < d_{q}^\mathrm{pre}(k_{q}) + t_{q}^\mathrm{trans} \le {d_{p}^\mathrm{pre}(k_{p})};\\
    0, & \mathrm{otherwise},
\end{cases}
\end{equation}
where $t_{q}^\mathrm{trans}$ represents the average transfer time from platform $q$ to the corresponding platforms at the same station.

After train service $k_p$ departs from platform $p$, the number of passengers waiting at the platform can be calculated by
\begin{equation}\label{lmpc-w}
n^\mathrm{after}_{p}(k_p) = n^\mathrm{before}_{p}(k_p) - n^\mathrm{depart}_{p}(k_p),
\end{equation}
where $n^\mathrm{after}_{p}(k_p)$ represents the number of passengers waiting at platform $p$ immediately after train service $k_p$ departs from platform $p$.



\section{Model predictive control for real-time train rescheduling}\label{lmpc-mpc_approach}
\subsection{Problem formulation}\label{lmpc-mpc_formualtion}
Based on the developed model, we can formulate the problem to minimize passenger delays and operational costs. The passenger delays corresponding to train service $k_p$ at platform $p$ can be formulated as
\begin{align}\label{lmpc-pass}
J_p^\mathrm{pass}(k_p) = n_{p}(k_p)\left( {d_{p}(k_p) - d^\mathrm{pre}_{p}(k_p)} \right) + n^\mathrm{after}_{p}(k_p)\left( { d^\mathrm{pre}_{p}(k_p+1) -  d_{p}(k_p)} \right).
\end{align}
As passengers expect to depart at the predetermined departure time $d^\mathrm{pre}_{p}(k_p)$, the term $n_{p}(k_p)\left( {d_{p}(k_p) - d^\mathrm{pre}_{p}(k_p)} \right)$ in (\ref{lmpc-pass}) represents the delay for passengers departing from platform $p$ with train service $k_p$; the term $n^\mathrm{after}_{p}(k_p)\left( { d^\mathrm{pre}_{p}(k_p+1)-  d_{p}(k_p)} \right)$ denotes the expected delay for passengers that could not board train service $k_p$, hence they have to wait for the next train at the platform.

Assigning more train units to a train service can increase the capacity for transporting passengers while leading to higher energy consumption. Furthermore, changing the train composition may require additional workload from operators and thus lead to additional costs. The operational costs of train service $k_p$ running from platform $p$ to its successor platform can be expressed as
\begin{equation}\label{lmpc-cost}
{J_p^\mathrm{cost}}(k_p) = \ell_{p}(k_p){E_p^\mathrm{energy}} + \eta_{k_p,p}{E_p^\mathrm{add}},
\end{equation}
where ${E_p^\mathrm{energy}}$ represents the average energy consumption for a train unit running from platform $p$ to its successor platform, $\ell_{p}(k_p){E_p^\mathrm{energy}}$ denotes the approximate energy consumption for $\ell_{p}(k_p)$ train units running from platform $p$ to its successor platform, and ${E_p^\mathrm{add}}$ denotes the additional cost for changing the train composition of a train service at platform $p$. 

In urban rail transit systems, a train service typically departs from a depot, visiting each platform along a line before returning to the depot. 
To ensure regular train departures at each platform, we define the time interval between two consecutive predetermined departure times as the control time step. The length of the control time step, denoted as $T_\mathrm{ctrl}$, is the same for all platforms along a line, and $\kappa$ is the index of the control time step\footnote{ $T_\mathrm{ctrl}$ is the same for different lines, but this scheme can be extended to cases where $T_\mathrm{ctrl}$ varies between lines.}.

Therefore, the optimization problem for the train rescheduling problem is
\begin{equation}\label{lmpc-problem}
\begin{array}{l}
\mathop {\min }\limits_{\scriptstyle{\bm{g}}(\kappa_0)}  \  J({\kappa_0}) := \sum\limits_{p \in \mathcal{P}} \sum\limits_{k_p \in \mathcal{N}_p(\kappa_0)} \left( {w_1 J_p^\mathrm{pass}(k_p) + w_2 J_p^\mathrm{cost}(k_p)} \right),\\
{\rm{s}}{\rm{.t}}{\rm{.}}\quad \mathrm{(\ref{lmpc-predetermined}) - (\ref{lmpc-cost})},
\end{array}
\end{equation}
where ${\bm{g}}({\kappa_0})$ represents a vector collecting all the variables of problem (\ref{lmpc-problem}), $\mathcal{P}$ is the set collecting all the platforms of the line, $\mathcal{N}_p(\kappa_0)$ is the set indices of trains departing from platform $p$ within the prediction time window starting at time step $\kappa_0$, and $w_1$ and $w_2$ are weights balancing two objectives.



\subsection{MINLP-based MPC for real-time train rescheduling}\label{lmpc-MINLP-MPC}
Problem (\ref{lmpc-problem}) contains piecewise constant (``if-then") constraints in (\ref{lmpc-y_kp}) and (\ref{lmpc-xi}). We apply the following transformation properties to convert (\ref{lmpc-y_kp}) into a mixed logical dynamical (MLD) system \citep{williams2013model}. 


{\bf{Transformation property 4.1}}:  If we introduce an auxiliary continuous variable $o_{k,p}$ and an auxiliary binary variable $\gamma_{k_p,p}$ with $\gamma_{k_p,p} = 1 \Leftrightarrow o_{k_p,p} = y_{p}(k_p)$ and $\gamma_{k_p,p} = 0 \Leftrightarrow o_{k_p,p} = -y_{p}(k_p)$, and then $o_{k_p,p} = |y_{p}(k_p)|$ is equivalent to
\begin{equation}\label{lmpc-trans_abs}
\left\{ {\begin{array}{*{40}{l}}
{o_{k_p,p} - y_{p}(k_p) \ge 0,}\\
{o_{k_p,p} - y_{p}(k_p) \le 2Y_\mathrm{max}(1-\gamma_{k_p,p}),}\\
{o_{k_p,p} + y_{p}(k_p) \ge 0,}\\
{o_{k_p,p} + y_{p}(k_p) \le 2Y_\mathrm{max}\gamma_{k_p,p},}\\
\end{array}} \right.
\end{equation}
where $Y_\mathrm{max}$ denotes the maximum value of $y_{p}(k_p)$. 

{\bf{Transformation property 4.2}}:
Based on \emph{Transformation property 4.1}, (\ref{lmpc-eta_kp}) is equivalent to $\eta_{k_p,p} = \left\{ {\begin{array}{*{20}{l}}
{1,\quad {\rm{if}}\quad {o_{k_p,p}} > 0;}\\
{0,\quad {\rm{otherwise}},}
\end{array}} \right.$, which can be
converted to
\begin{equation}\label{lmpc-trans_gamma}
\left\{ {\begin{array}{*{30}{l}}
{o_{k_p,p} \le \eta_{k_p,p}O_\mathrm{max},}\\
{o_{k_p,p} \ge \epsilon + (1 - \eta_{k_p,p})(O_\mathrm{min} - \epsilon).}
\end{array}} \right.
\end{equation}
where $O_\mathrm{max}$ and $O_\mathrm{min}$ are the minimum and maximum values of $o_{k_p,p}$, respectively, and $\epsilon$ is a sufficiently small number, typically representing machine precision. 

Therefore, (\ref{lmpc-y_kp}) can be transformed to
\begin{equation}\label{lmpc-y_kp_linear}
\tau^\mathrm{add}_{p}(k_p) = \eta_{k_p,p} t^\mathrm{cons}_{p}.
\end{equation}

{\bf{Transformation property 4.3}}: 
If we define $m_\mathrm{a}$ and $M_\mathrm{a}$ as the minimum and maximum values of $a_{p'}(k_{p'})$, respectively, then following the transformation property in \cite{bemporad1999control}, (\ref{lmpc-xi}) is equivalent to the following inequalities
\begin{equation}\label{lmpc-transif}
\left\{ {\begin{array}{*{20}{l}}
{{a_{p'}(k_{p'})} - {a_{p}(k_{p})} \le \left( {1 - \xi_{k_{p'},k_p,p',p}} \right)\left( {{M_{\rm{a}}}  \!-\! {a_{p}(k_{p})}} \right),}\\
{{a_{p'}(k_{p'})} - {a_{p}(k_{p})} \ge \varepsilon  + {\xi_{k_p,k_{p'},p,p'}}\left( {{m_{\rm{a}}} \!-\! a_{p}(k_{p}) \!-\! \varepsilon } \right).}
\end{array}} \right.
\end{equation}

Based on the transformations described above, we can convert problem (\ref{lmpc-problem}) into an MINLP problem. The nonlinearity in the MINLP arises from the nonlinear objective function. The integer variables in this problem encompass the train composition variables ($y_{p}(k_p)$ and $\ell_{p}(k_p)$), the train ordering variable ($\xi_{k_p,k_{p'},p,p'}$), and auxiliary binary variables ($\gamma_{k_p,p}$ and $\eta_{k_p,p}$). 

For compactness, we rewrite the resulting MINLP problem in the following form:
\begin{subequations}\label{lmpc-pro}
\begin{align}
&{\mathop {\min }\limits_{\scriptstyle{\bm{x}}(\kappa_0),{\bm{u}}(\kappa_0),{\bm{\delta}}(\kappa_0)} \!  J(\kappa_0) := \sum\limits_{\kappa = {\kappa_0}}^{{\kappa_0}+ {N} - 1} {L(x(\kappa),u(\kappa),\delta(\kappa))} } \label{lmpc-obj}\\
&\quad {\rm{s}}.{\rm{t}}.\quad {x}(\kappa + 1) = {A_\kappa}{x}(\kappa) + {B_{1,\kappa}}{u}(\kappa) + {B_{2,\kappa}}{\delta }(\kappa),\label{lmpc-state}\\
& \quad \qquad {D_{3,\kappa}}{x}(\kappa) + {D_{1,\kappa}}{u}(\kappa) +  {D_{2,\kappa}}{\delta}(\kappa)  \le  {D_{4,\kappa}},\label{lmpc-cons}\\
& \quad  \qquad \kappa = {\kappa_0}, \cdots ,{\kappa_0} + N - 1, \nonumber
\end{align}
\end{subequations}
where $N$ is the total number of time steps, ${x}(\kappa)$, ${u}(\kappa)$, and ${\delta}(\kappa)$ collect all the independent variables, continuous decision variables, and discrete decision variables for time step $\kappa$, respectively, and ${\bm{x}}(\kappa_0) = [{x}^\intercal (\kappa_0), {x}^\intercal(\kappa_0+1), \ldots, {x}^\intercal(\kappa_0+N-1)]^\intercal $, ${\bm{u}}(\kappa_0) = [{u}^\intercal (\kappa_0), {u}^\intercal(\kappa_0+1), \ldots, {u}^\intercal(\kappa_0+N-1)]^\intercal$, and ${\bm{\delta}}(\kappa_0) = [{\delta}^\intercal (\kappa_0), {\delta}^\intercal(\kappa_0+1), \ldots, {\delta}^\intercal(\kappa_0+N-1)]^\intercal$. In (\ref{lmpc-pro}),  ${L(x(\kappa),u(\kappa),\delta(\kappa))}$ represents the nonlinear objective function for time step $\kappa$, (\ref{lmpc-state}) collects all equality constraints, and (\ref{lmpc-cons}) collects all inequality constraints.

\begin{figure}[htbp]
\begin{center}
\includegraphics[width=6cm]{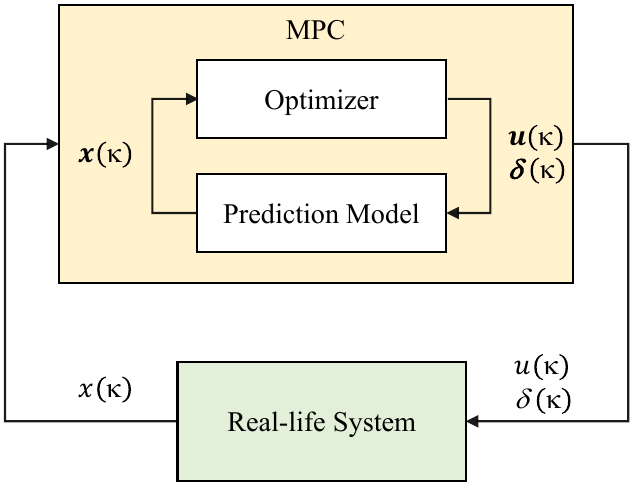}    
\caption{Model predictive control for real-time train rescheduling.}
\label{lmpc-mpc}
\end{center}
\end{figure}

Solving (\ref{lmpc-pro}) leads to a series of continuous decision variables and discrete decision variables, and only the decision variables at time step $\kappa_0$ are applied. At the next time step, the prediction time window is shifted for one step, and a new optimization is formulated. The framework is depicted in Fig.~\ref{lmpc-mpc}.

{\bf{Lemma 1}} (\emph{Recursive Feasibility}): If problem (\ref{lmpc-pro}) is feasible at time step $\kappa$ with initial state $x(\kappa)$, then problem (\ref{lmpc-pro}) is also feasible at time step $\kappa+1$. 
\begin{proof}
    The proof relies on finding a feasible solution for time step $\kappa+1$. Recall that the planning time window at each platform is divided into several intervals of equal length, with a train service departing from the platform at each interval according to (\ref{lmpc-predetermined}). In general, there are two types of depots: (i) depots connected to the terminal platform (e.g., Platform 1 in Fig.~\ref{lmpc-line}), and (ii) depots connected to intermediate platforms (e.g., Platform 2 in Fig.~\ref{lmpc-line}).
    
    i) For a depot connected to the terminal platform: if the problem (\ref{lmpc-pro}) is feasible at time step $\kappa$, then at each time step, a train service returns to the depot from the opposite direction of the line. In this context, the new train service departing from the terminal platform at step $\kappa + 1$ can directly utilize the train units by performing a turnaround from the opposite direction of the line following (\ref{lmpc-turnaround}). Moreover, utilizing one train unit when performing the turnaround action is always feasible.

    ii) For any depot connected to an intermediate platform: a feasible solution is obtained by maintaining the composition of each train service the same as it was at time step $\kappa$. 
\end{proof}

\subsection{MILP-based MPC for real-time train rescheduling}\label{lmpc-MILP-MPC}
In Section~\ref{lmpc-mpc_formualtion}, a nonlinear objective function has been defined in (\ref{lmpc-pass}) to calculate the passenger delays. This nonlinear objective function results in the MINLP-based MPC approach formulated in Section~\ref{lmpc-MINLP-MPC}. In general, the nonlinear term significantly increases the computational burden. In what follows, we simplify the nonlinear objective function to reduce the computational burden. 

Since we divide the planning time window as indicated in Fig.~\ref{lmpc-demands_original} and the actual departure time $ d_{p}(k_p)$ is constrained by
$d^\mathrm{pre}_{p}(k_p) \le d_{p}(k_p) < d^\mathrm{pre}_{p}(k_p+1)$, by using the upper bound and lower bound of $d_{p}(k_p)$ we approximate the nonlinear objective function (\ref{lmpc-pass}) by 
\begin{align}\label{lmpc-pass_linear}
J_p^\mathrm{pass}(k_p) \approx w_3 n_{p}(k_p)\left( {d^\mathrm{pre}_{p}(k_p+1) - d^\mathrm{pre}_{p}(k_p)} \right) + n^\mathrm{after}_{p}(k_p)\left( { d^\mathrm{pre}_{p}(k_p+1) -  d^\mathrm{pre}_{p}(k_p)} \right),
\end{align}
where $w_3$ is a weight used to balance the approximated errors. In particular, $w_3$ can be defined as
\begin{equation}\label{lmpc-w3}
    w_3 = \frac{1}{ \sum_{p\in \mathcal{P}}|\mathcal{I}_p|}\sum_{p\in \mathcal{P}}\sum_{k_p\in \mathcal{I}_p}\frac{\bar d_{p}(k_p) -  d^\mathrm{pre}_{p}(k_p)}{d^\mathrm{pre}_{p}(k_p+1) -  \bar d_{p}(k_p)},
\end{equation}
where $|\mathcal{I}_p|$ represents the cardinality of $\mathcal{I}_p$, and $ \bar d_{p}(k_p)$ represents the average value of  $ d_{p}(k_p)$ from historical data.

Other settings are identical to those of Section~\ref{lmpc-MINLP-MPC}. With this approximation, a linear objective function (\ref{lmpc-pass_linear}) is obtained, and due to the objective function and constraints being all linear, we obtain an MILP-based MPC approach for real-time train rescheduling.

\section{Learning-based MPC for real-time train rescheduling}\label{lmpc-lmpc_approach}
For the MINLP-based MPC and MILP-based MPC in Section~\ref{lmpc-mpc_approach}, an MINLP or MILP problem should be solved at each step, which is typically not computationally affordable for real-time application as the number of integer variables significantly influences the computational complexity. 
\cite{cauligi2022prism} developed a presolve and recurrent network-based mixed-integer solution method (PRISM) to handle mixed-integer convex programming problems, where an LSTM network generates the integer variables, and then a convex optimization problem is solved to improve the solution efficiency. 
To handle computational complexity issues arising in train rescheduling problems, based on PRISM \citep{cauligi2022prism}, we develop a learning-based MPC approach with presolve techniques tailored to the train rescheduling problem with flexible train composition. In this setting, recurrent neural networks, such as LSTM neural networks, and presolve techniques are used to assign the integer variables of an MPC problem.
As a result, at each time step, the MPC optimizer only needs to solve a continuous nonlinear optimization problem with fewer variables than the original problem. 
The main difference between the proposed approach and that in \cite{cauligi2022prism} consists of the presolve techniques and integration with the MPC framework, which are specifically designed for real-time train rescheduling.


\subsection{Presolve techniques}\label{lmpc-presolve}
Presolve techniques streamline optimization processes by pruning a subset of decision variables with values predetermined by other coupled variables and constraints, setting them to predefined values. In this paper, we develop the following presolve techniques.  


{\bf{Presolve technique 5.1:}} As we adjust the departure time of train service $k$ at platform $p$ according to $d^\mathrm{pre}_{p}(k_p) \le d_{p}(k_p) < d^\mathrm{pre}_{p}(k_p+1)$, the departure order between some trains has already been determined: If $d^\mathrm{pre}_{p}(k_p) \ge d^\mathrm{pre}_{p'}(k_{p'}+1) + t_{p'}^\mathrm{roll}$, then $\xi_{k_p,k_{p'},p,p'} = 1$. If $d^\mathrm{pre}_{p}(k_p+1) \le d^\mathrm{pre}_{p'}(k_{p'}) + t_{p'}^\mathrm{roll}$, then $\xi_{k_p,k_{p'},p,p'} = 0$. 

{\bf{Presolve technique 5.2:}} According to the definition of $\xi_{k_p,k_{p'},p,p'}$ in (\ref{lmpc-xi}), the order of trains at the same platform should be kept consistent, i.e., $\xi_{k_p+1,k_{p'},p, p'} \ge \xi_{k_p,k_{p'},p,p'}$. 

{\bf{Presolve technique 5.3:}} The train composition cannot be changed at a station that is not linked with a depot: If $\sigma_p = 0$, then $y_{p}(k_p) = 0$.

{\bf{Presolve technique 5.4:}} Let $t_0 = \kappa_0 T$ represent the current time. If train service $k_p$ has already departed from station $p$ at time $t_0$, the composition cannot be changed at that station: If $d_{p}(k_p) \le t_0$, $y_{p}(k_p) = y_{p}^*(k_p), \ \forall p \in \mathcal{P}$, where $y_{p}^*(k_p)$ represents the value of $y_{p}(k_p)$ obtained before train $k$ departs from platform $p$ and $\mathcal{P}$ denotes the set of all platforms of the line.

\subsection{Environment setting}
The environment of the learning-based MPC algorithm includes the system and an MPC optimization problem. The state and variables that interact with the environment are defined as follows: 

{\bf{State}} $\bm{s}(\kappa) \in S$: The state space ought to encompass all necessary information of the framework so that the neural network can be trained such that the input can capture as much possible situations as possible. Hence, state at time step $k$ is defined as:
\begin{equation}
    \bm{s}(\kappa) = [\bm{n}^\intercal(\kappa), \bm{\rho}^\intercal(\kappa), \bm{N}^\intercal(\kappa) ]^\intercal,
\end{equation}
where $\bm{n}(\kappa)$ includes the variables $n_{p}(k_p)$ for train service $k_p$ at its corresponding platform $p$ with $\kappa T \in [d_{p}^\mathrm{pre}(k_p-1), d_{p}^\mathrm{pre}(k_p))$,  $\bm{\rho}(\kappa)$ collects the passenger demands $\rho_{p}(k_p)$ for all train services $k_p$ departing from all platforms $p$ from time $t = \kappa T$ to the end of the prediction time window, and $\bm{N}(\kappa)$ collects the number of available trains for all depots at time  $t = \kappa T$.

{\bf{Discrete variables}} $\bm{\delta}(\kappa) \in A$: The discrete variables correspond to the discrete variables of the MPC optimization problem (\ref{lmpc-pro}) in each step. Before we evaluate the discrete variable, we first implement the presolve techniques in Section~\ref{lmpc-presolve} to avoid infeasible actions, and then, we solve the resulting problem corresponding to the discrete action. 

{\bf{Continuous variables}} $\bm{u}(\kappa) \in U$: The continuous variables represent the continuous decision variables at time step $\kappa$ in the MPC optimization problem (\ref{lmpc-pro}).

\subsection{Offline training for learning-based algorithms}\label{lmpc-offline}
In practice, the train schedules across consecutive time intervals are interdependent 
due to the headway relation between trains and the physical connections between stations. The neural network should be able to capture and retain essential information over sequences time intervals. 
Therefore, a long short-term memory (LSTM) network \citep{hochreiter1997long,cauligi2022prism} is applied to train the agent. As a deep recurrent neural network (RNN), the LSTM architecture enables the network to remember the dynamic interdependencies within train schedules, ensuring effective adaptation and learning in response to evolving temporal dynamics.


\begin{figure}[htbp]
\begin{center}
\includegraphics[width=9cm]{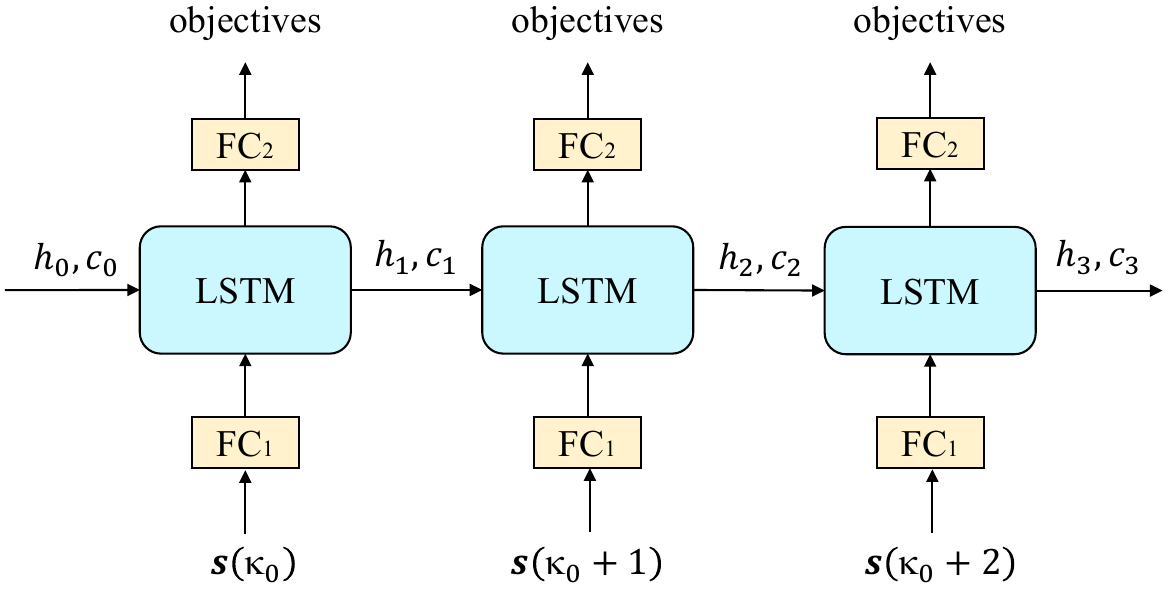}    
\caption{Training procedure of the learning-based MPC approach with LSTM.}
\label{lmpc-LSTM}
\end{center}
\end{figure}

An ensemble of LSTM networks is trained based on the MINLP approaches to improve the solution's feasibility rate. For each network, the training procedure for the learning-based approach is shown in Fig.~\ref{lmpc-LSTM}, where $h_{\kappa}$ represents the hidden state vector, $c_{\kappa}$ denotes the cell state vector, and FC$_1$ and FC$_2$ are feedforward blocks. The block FC$_1$ is a fully connected layer with an identity activation function that transforms the state vector into a vector with the size of the hidden state. 
The block FC$_2$ is a fully connected layer with a softmax activation function, and the resulting output represents the likelihood of possible action sequences. 
At each training step, a random day from the dataset is selected, followed by a random time step from the timetable. The LSTM network takes the current state and the hidden state as inputs 
to generate the objective function values for possible discrete variables. 

The mean squared error (MSE) is applied to update the parameters of the LSTM network as the optimizer with the following loss function:
\begin{equation}\label{lmpc-mse}
L^\mathrm{cross} = \frac{1}{|\mathcal{S}|}\sum_{\bm{s} \in \mathcal{S}}\sum_{\kappa = \kappa_0}^{\kappa_0 + N-1} \big(J_{\bm{s}}^*(\kappa) -  J_{\bm{s}}(\kappa,\bm{\delta}(\kappa)) \big)^2,
\end{equation}
where $L^\mathrm{cross}$ represents the loss function, 
$\mathcal{S}$ defines a set collecting all transitions, $|\mathcal{S}|$ denotes the cardinality of $\mathcal{S}$,  $J^*_{\bm{s}}(\kappa)$ represents the optimal objective function value for the state $\bm{s}$ at time step $\kappa$, obtained by solving the resulting MINLP problem, and $J_{\bm{s}}(\kappa, \bm{\delta}(\kappa))$ represents the objective function value at time step $\kappa$, corresponding to the state transition $\bm{s}$ and the discrete variable $\bm{\delta}(\kappa)$, generated by the LSTM network, i.e., objectives in Fig.~\ref{lmpc-LSTM},. 

At the next training step, the hidden states $h_{\kappa}$ and $c_{\kappa}$ from the current step are passed to the LSTM network. A new data set group is then randomly selected, followed by a random time step selection. Then, (\ref{lmpc-mse}) is applied to update the LSTM network parameters. This procedure is repeated until the end of the training process.




\subsection{Online implementation of learning-based MPC}

\begin{figure}[htbp]
\begin{center}
\includegraphics[width=8cm]{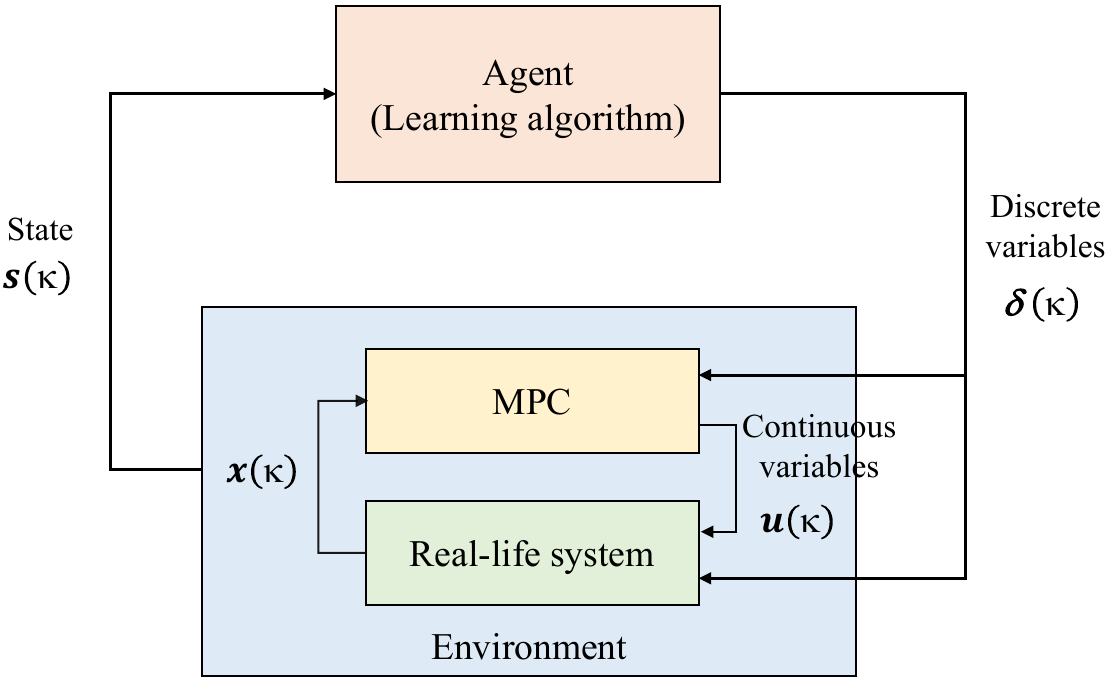}    
\caption{Learning-based MPC for real-time train rescheduling.}
\label{lmpc-SL_MPC}
\end{center}
\end{figure}

The framework of the learning-based MPC approach is shown in Fig.~\ref{lmpc-SL_MPC}. At each step, pruned discrete variables are generated based on the current state and the neural networks trained in Section\ref{lmpc-offline}. These pruned discrete variables, combined with the presolve techniques from Section~\ref{lmpc-presolve}, allow for the determination of all discrete variables. Once the discrete variables are established, the MPC optimization problem from Section~\ref{lmpc-MINLP-MPC} becomes a continuous-variable nonlinear programming (NLP) problem, while the problem in Section~\ref{lmpc-MILP-MPC} simplifies to a linear programming (LP) problem. Solving the resulting NLP or LP problem yields the optimal continuous variable values, which are used to generate the timetable. After implementing the timetable, the new state of the urban rail transit network is observed and used to generate pruned discrete variables for the next time step. This procedure is repeated until the control process is complete. 

To improve the feasibility rate of the integer variables generated by LSTM networks, for each branch of LSTM networks trained in Section~\ref{lmpc-offline}, the networks are sequentially employed to generate integer variables. 
Moreover, {\emph{Lemma 1}} can be applied when LSTM networks fail to generate feasible integer variables, ensuring that feasible integer variables are always obtained. 
In this context, the recursive feasibility of the learning-based MPC approach can be guaranteed.

\section{Case study}\label{lmpc-section6}
\subsection{Basic setting}
\begin{figure}[htbp]
\begin{center}
\includegraphics[width=9cm]{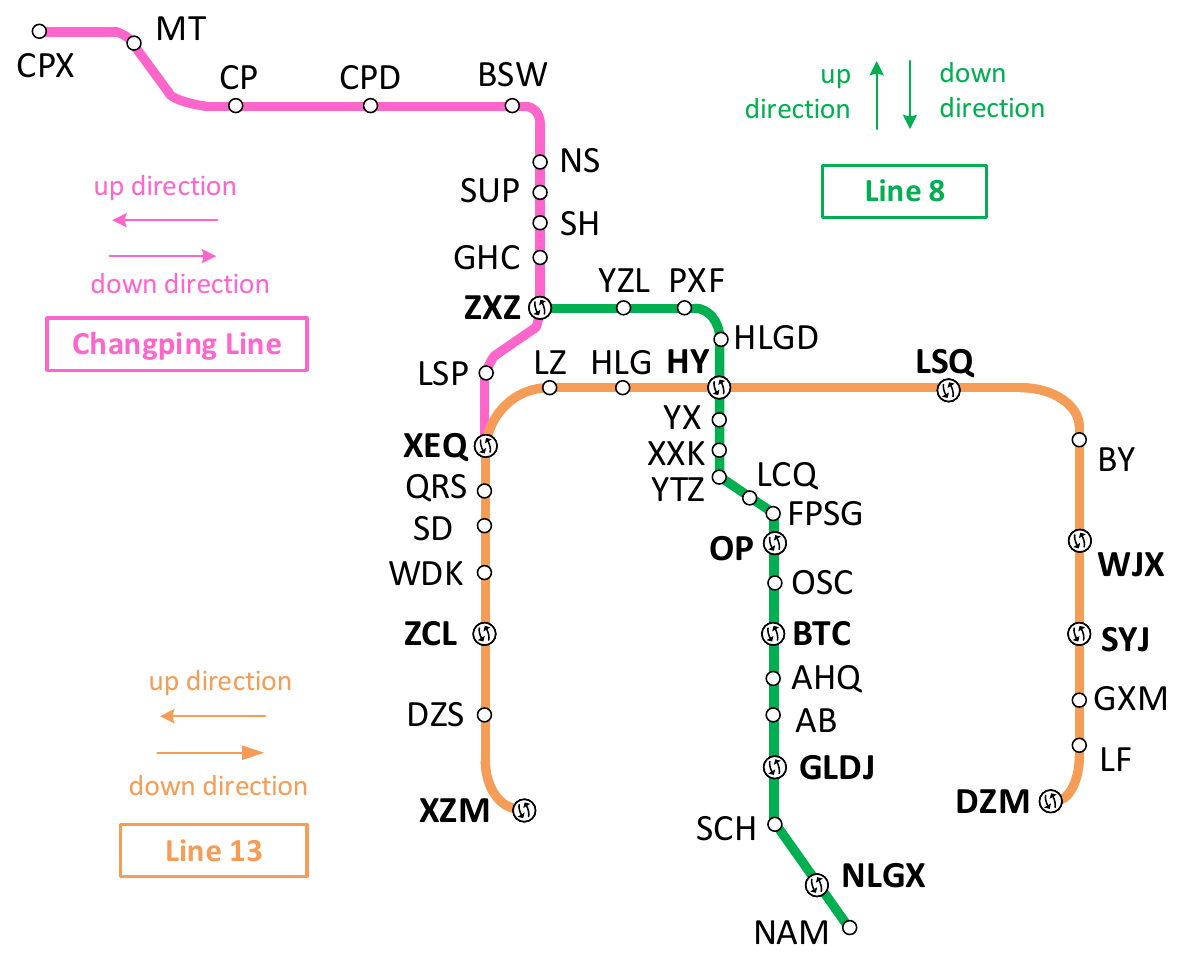}    
\caption{The layout of considered urban rail transit network (with 3 lines).}
\label{lmpc-3lines}
\end{center}
\end{figure}

In this section, we illustrate the proposed approaches based on real-life data from a network of three lines in the Beijing urban rail transit network. As shown in Fig.~\ref{lmpc-3lines}, the network including 3 bi-directional lines with 45 stations. There are 3 transfer stations, i.e., Station ZXZ, Station XEQ, and Station HY, where passengers can transfer from one line to another.
For each line, there is a depot connected with the starting station of the line, i.e., Station CPX for Changping Line, Station XZM for Line 13, and Station ZXZ for Line 8. 
The values of parameters for the case study are given in Table~\ref{lmpc-parameters}.   The original timetable is generated based on the regular headway, regular dwell time, and average running times in Table~\ref{lmpc-parameters}.  According to the definition, the length of a time step is the sum of the regular headway and the regular dwell time. The number of train units in the depot for each line has been selected as a random integer number with the value varying among the range given in Table~\ref{lmpc-parameters}. 

The length between every two consecutive stations is openly accessible on the website of Beijing Subway\footnote{\url{https://www.bjsubway.com/station/zjgls/}}.  In the case study, the average running time and the average energy consumption of a train between every two consecutive stations are calculated using the method in \cite{wang2015efficient} with the maximum acceleration of 0.75 m/s$^2$, the maximum deceleration of 0.7 m/s$^2$, and the cruising speed of 70 km/h, respectively. The sectional passenger demands are obtained based on real-life passenger flow data from the Beijing urban rail transit network, collected in January 2020. We have selected data from 6:00 AM to 10:00 PM for simulation, so the data contains both peak hours and off-peak hours. For training agents and simulation, we have generated passenger demands based on a Poisson distribution by using real-life passenger flow data as the expected value.   

\begin{table}[htbp]
\centering
\caption{Main parameters for the case study}\label{lmpc-parameters}
\begin{tabular}{lll} 
\hline
Parameter & Symbol & Value \\ \hline
Regular headway & $h_p^\mathrm{regular}$ & 180 s \\
Minimum headway & $h_p^\mathrm{min}$ &120 s\\
Regular dwell time& $\tau_p^\mathrm{regular}$ & 60 s\\
Minimum dwell time & $\tau_p^\mathrm{min}$ &  30 s \\
Average turnaround time & $r_p^\mathrm{turn}$ & 52.9 s \\
Minimum running time & $r_p^\mathrm{min}$ &  $0.8 \cdot r_p^\mathrm{avrg}$ \\
Maximum running time & $r_p^\mathrm{max}$ &  $1.2 \cdot r_p^\mathrm{avrg}$ \\
Time for changing train composition & $t_p^\mathrm{cons}$ & 60 s \\
Time for rolling stock circulation & $t_p^\mathrm{roll}$ & 240 s \\
Transfer rate at a transfer station&  $\beta_{q,p}$ & 10\% \\
Capacity of a train unit & $C_\mathrm{max}$ & 400 persons \\
Regular train composition of a train service& $\ell_p^\mathrm{regular}$ & 2 train units \\
\makecell[l]{Minimum number of train units\\ \ \ included in a train service} & $\ell_p^\mathrm{min}$ &  1 train unit\\
\makecell[l]{Maximum number of train units\\ \ \ included in a train service} & $\ell_p^\mathrm{max}$ &  4 train units\\
Weighted term  &  $w_1$   &  $10^{-4}$\\
Weighted term  &  $w_2$   &  $10^{-1}$\\
Weighted term  &  $w_3$ &  $10^{-1}$\\
Number of train units for Changping Line& $N_z^{\mathrm{train}}$ & [55, 75]\\
Number of train units for Line 13& $N_z^{\mathrm{train}}$ & [70, 90]\\
Number of train units for Line 8& $N_z^{\mathrm{train}}$ & [60, 80]\\
Prediction horizon of MPC & $N$ & 40 \\
\hline
\end{tabular}
\end{table}

The simulations have been conducted using Python as a programming language, PyTorch as the machine learning library, and \texttt{gurobi} to solve optimization problems.  Adam \citep{kingma2014adam} is applied in the offline training process to minimize MSE, and dropout \citep{srivastava2014dropout} is used to handle the over-fitting issue.  Moreover, the experiments were conducted on a computing cluster with Intel XEON E5-6248R CPUs. The dataset consists of $96000$ states and the corresponding optimal solutions, and it has been built using 60 CPU cores and 240GB of RAM in $24$ hours. The training and hyperparameter tuning processes have been conducted using 144 CPU cores, 864GB of RAM, using more than 200000 iterations for $24$ hours. The code used to generate the results in this paper is available online \footnote{\url{https://github.com/fabcaio/railwaynet\_learning/}}.

To reduce the solution time of the MINLP solver without significantly compromising optimality, a simple early termination criterion was employed: if the optimality gap does not decrease by 0.5\% within 10 seconds, the solution process terminates, and the solver outputs the best solution found. Moreover, as the MILP-based approach typically has a significantly shorter solution time than the MINLP-based approach, the integer variables generated by the MILP-based approach are used as a warm-start rule for the MINLP-based approach. 

In this section, the developed train rescheduling approaches, i.e., MINLP-based MPC, warm-start-MINLP-based MPC, MILP-based MPC, learning-NLP-based MPC, and learning-LP-based MPC are evaluated. 
As defined in Section~\ref{lmpc-section3}, the length of a time step is 240\ s. Hence, to ensure that a solution can be obtained for each time step, we set the maximum solution time for each approach as 240\ s. In addition, as a longer solution time typically yields better objective function value, we use the MINLP approach with warm-start and a longer maximum solution time, i.e., 600\ s, as a benchmark to evaluate the performance of the developed approaches.  

To improve the overall feasibility of the learning-based approaches, 15 LSTM networks were trained separately, and the 15 LSTM networks were sequentially employed for each branch to generate integer variables. The inference process of the LSTM networks in the ensemble is performed sequentially, where the $(i\!+\!1)$th network is evaluated only if the $i$th and all preceding networks fail to produce a feasible solution.  
In particular, the ensemble consists of LSTM networks with hidden sizes from the set $\{512,\ 1024\}$, dropout rates from the set $\{0,\ 0.5\}$, learning adjusting in the set $\{\mathrm{on},\ \mathrm{off}\}$, and output masking in the set $\{\mathrm{on},\ \mathrm{off}\}$.  

\subsection{Simulation and results}

\begin{figure}[htbp]
\begin{center}
\includegraphics[width=8.5cm]{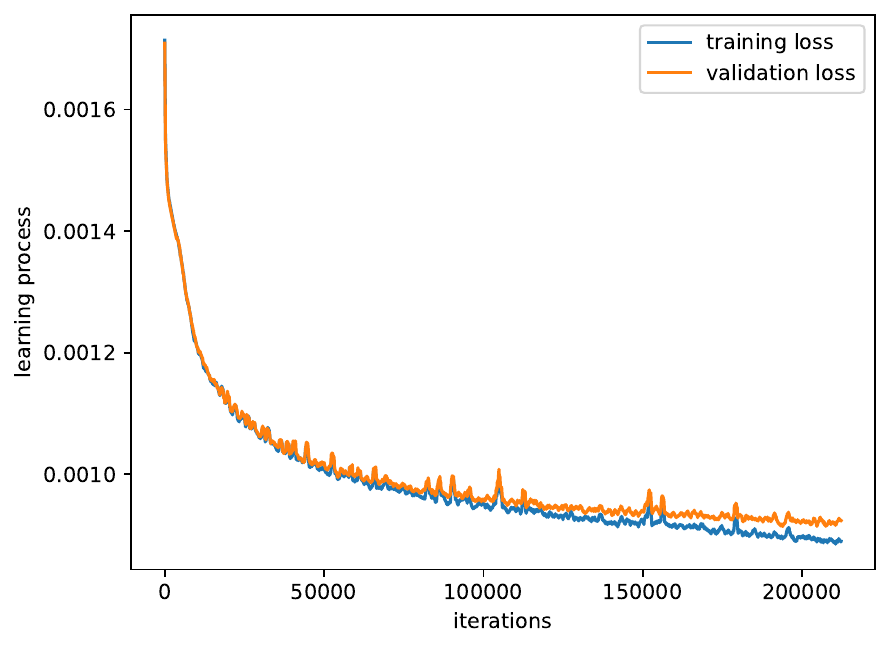}    
\caption{Learning process of the learning algorithm.}
\label{lmpc-learning_process}
\end{center}
\end{figure}

\begin{table*}[htbp]
\centering
\caption{Open-loop optimization for real-time train rescheduling}\label{lmpc-open-loop}
\begin{tabular}{c|c|ccc|ccc|c} \hline
\multirow{2}*{Approach}& \multirow{2}*{Warm-start}& \multicolumn{3}{c}{Optimality gap}  & \multicolumn{3}{c}{CPU time (s)} & \multirow{2}*{Feasibility rate}   \\ \cline{3-8}
&  & max & average& min & max  & average& min   & \\ \hline
Benchmark & yes&- & -&- &  600.10 &222.18 & 3.67 &  100\% \\ 
MINLP &no& 100.81\% & 7.22\% &0\%  &240.14 & 239.84 &52.47  & 100\% \\ 
Warm-start MINLP & yes  & 12.96\% &  0.04\% & -0.24\% &240.10 & 96.71 & 3.58 & 100\% \\ 
MILP & no &1.54\% & 0.47\%  & -33.73\% & 240.01&  8.77 &0.37&  100\% \\ 
Learning + NLP & no  &3.48\%&-0.11\%  &-34.28\%  &112.22 &6.89 &1.80 & 98.94\% \\ 
Learning + LP& no &1.73\% &0.22\%  &-33.42\%  &0.25 & 0.13&0.11 & 98.55\%  \\ 
\hline
\end{tabular}
\end{table*}
We have conducted simulations to train the learning algorithm, and the learning process for one of the networks is shown in Fig.~\ref{lmpc-learning_process}, where a 1000-step moving average approach is applied to smooth the learning curve. From Fig.~\ref{lmpc-learning_process}, we can see that the learning curve descends very quickly during the first 100000 iterations, and then the performance gradually improves until iteration 200000. 

We perform simulations for both open-loop optimization, where the MPC optimization problem is solved for one step, and closed-loop MPC, involving real-time MPC optimization over 30 steps. For open-loop optimization, we have performed simulations using the developed learning-based approaches on 10 separate cores for 24 hours  (resulting in 1054 scenarios). For comparison, simulations have also been performed with the Benchmark approach, the MINLP approach, the warm-start MINLP approach, and the MILP approach, where the Benchmark approach corresponds to the MINLP approach with warm-start and a maximum solution time of 600\ s. 

The optimality gap, CPU time, and feasibility rate among the simulations are given in Table~\ref{lmpc-open-loop}.  From Table~\ref{lmpc-open-loop}, we see that the MINLP approach without warm-start has the worst optimality gap and the worst CPU times. In general, the nonlinear objective function and the presence of integer decision variables significantly influence the performance of MINLP. By applying the warm start, the optimality gap and the solution time are reduced; however, the maximum solution time and the average solution time of the warm-start MINLP approach are still large with values of 240.10\ s and 96.71\ s, respectively.  By approximating the nonlinear objective function, the solution time of the MILP approach is further reduced without sacrificing too much optimality. However, the MILP also reached the maximum allowed solution time in some cases, i.e., the maximum CUP time of MILP in Table~\ref{lmpc-open-loop} still reached 240\ s. 

In the open-loop optimization, the developed learning-based NLP approach (Learning + NLP) and learning-based LP approach (Learning + LP) achieve comparable performance with an average optimality gap of -0.11\% and 0.22\% among the feasible cases, while significantly reducing the solution time to an average of 6.89\ s and 0.13\ s, respectively.
Furthermore, by approximating the nonlinear objective function, the learning-based LP approach further reduces the solution time, enabling the optimized timetable to be obtained in under 1 second. Both the learning-based NLP and LP approaches demonstrate high feasibility rates, at 98.94\% and 98.55\%, respectively.  




To further demonstrate the performance of the learning-based MPC approach, we conduct closed-loop control tests on 10 separate cores over 24 hours, with each episode consisting of 30 control steps. If the integer variables generated by the learning agents are infeasible, the heuristic described in \emph{Lemma 1} is applied to obtain the train decomposition and to generate a feasible timetable. 
The simulation results are presented in Table~\ref{lmpc-closed-loop}. It can be observed that after applying the heuristic described in \emph{Lemma 1}, the feasibility rate reached 100\%. Compared to the warm-start MINLP and MILP approaches, the optimality of learning-based approaches is reduced in some cases because heuristic rules typically generate suboptimal solutions. In the closed-loop case, the maximum solution time for traditional approaches, i.e., warm-start MINLP and MILP, reaches 240 s, making them unsuitable for real-life applications where operators require an optimized solution as soon as possible for real-time timetable rescheduling. In contrast, the solution times for learning-based NLP and LP approaches are significantly lower, with average solution times of 6.50 s and 0.13 s, respectively. Table~\ref{lmpc-closed-loop} indicates that learning-based approaches significantly enhance solution efficiency, with an acceptable optimality sacrifice of approximately 5\%.




\begin{table*}[htbp]
\centering
\caption{Closed-loop MPC for real-time train rescheduling}\label{lmpc-closed-loop}
\begin{tabular}{c|c|ccc|ccc|c} \hline
\multirow{2}*{Approach}& \multirow{2}*{Warm-start}& \multicolumn{3}{c}{Optimality gap}  & \multicolumn{3}{c}{CPU time (s)} & \multirow{2}*{Feasibility rate}   \\ \cline{3-8}
&  & max & mean& min & max  & mean& min   & \\ \hline
Benchmark & yes&- & -&- &  600.12 &227.18 & 0.05 &  100\% \\ 
Warm-start MINLP & yes  & 13.03\% &  -0.22\% & -60.95\% &240.06 & 51.36& 0.05 & 100\% \\ 
MILP & no &21.18\% & 0.46\%  & -35.61\% & 240.05&  6.83 &0.06&  100\% \\ 
Learning + NLP & no  &93.43\%&4.92\%  &-11.02\%  &91.20 &6.50 &1.84 & 100\% \\ 
Learning + LP& no &126.71\% &5.50\%  &-9.99\%  &0.31 & 0.15&0.11 & 100\%  \\ 
\hline
\end{tabular}
\end{table*}

\begin{figure}[htbp]
\begin{center}
\includegraphics[width=9cm]{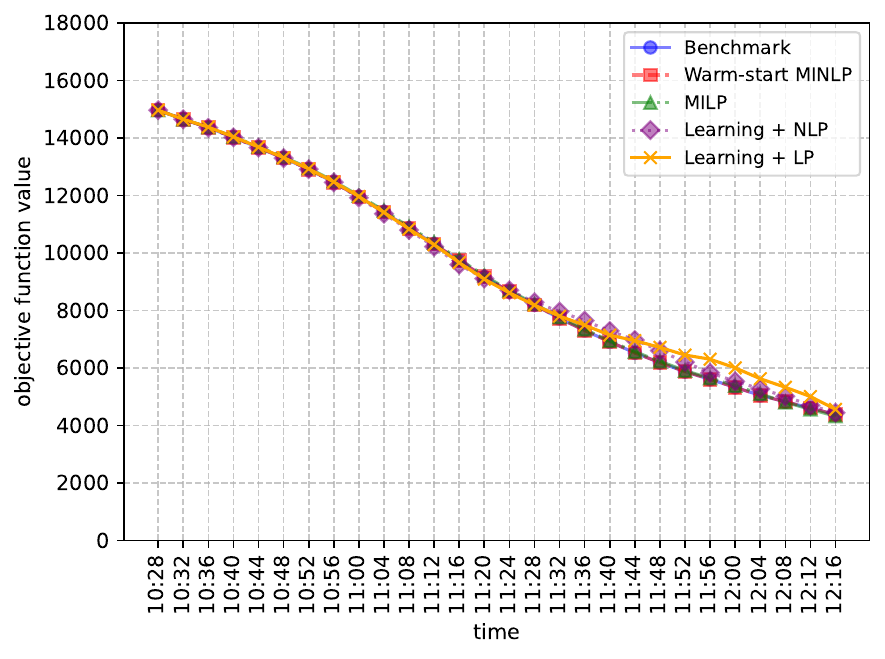}    
\caption{Objective function value at each MPC step for one episode.}
\label{lmpc-MPC_cost}
\end{center}
\end{figure}

\begin{figure}[htbp]
\begin{center}
\includegraphics[width=8.5cm]{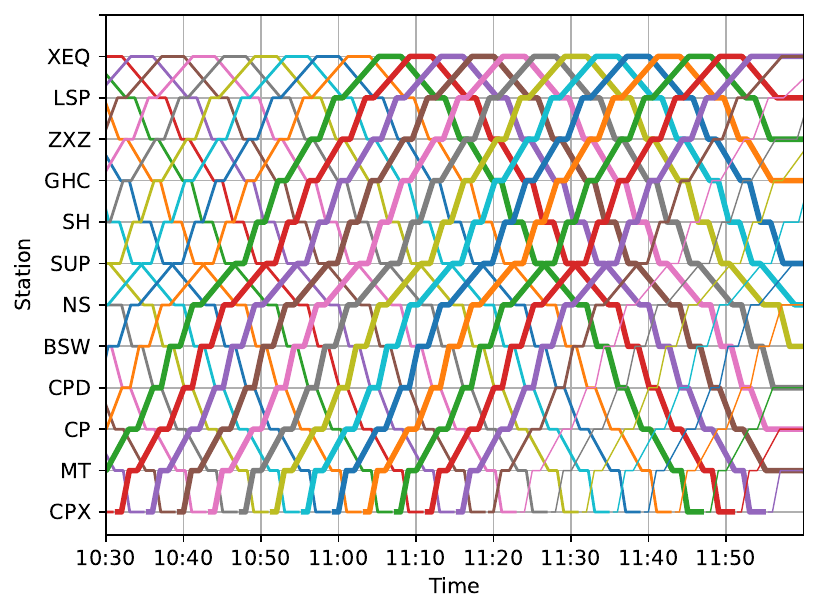}    
\caption{Timetable for Changping Line of the benchmark approach} 
\label{lmpc-benchmark_timetable}
\end{center}
\end{figure}
\begin{figure}[htbp]
\begin{center}
\includegraphics[width=8.5cm]{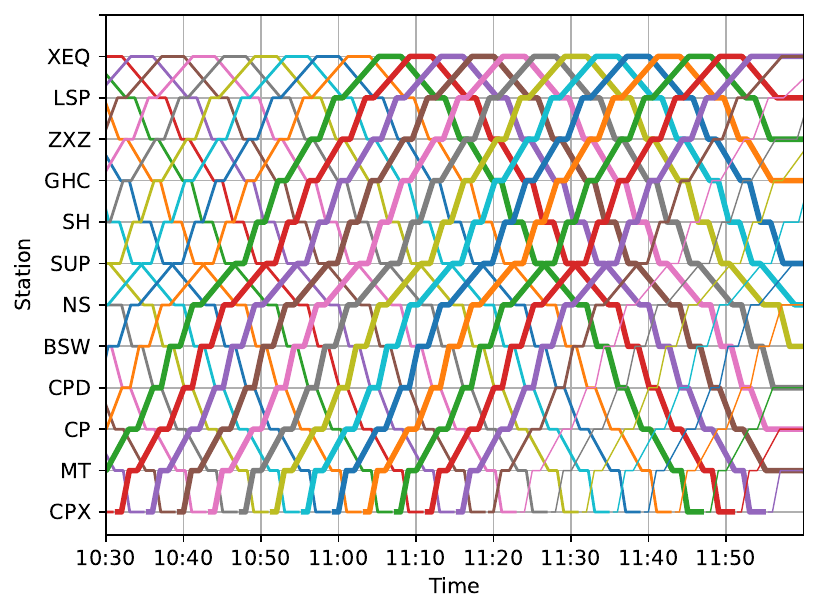}    
\caption{Timetable for Changping Line of the warm-start MINLP approach} 
\label{lmpc-minlp_timetable}
\end{center}
\end{figure}
\begin{figure}[htbp]
\begin{center}
\includegraphics[width=8.5cm]{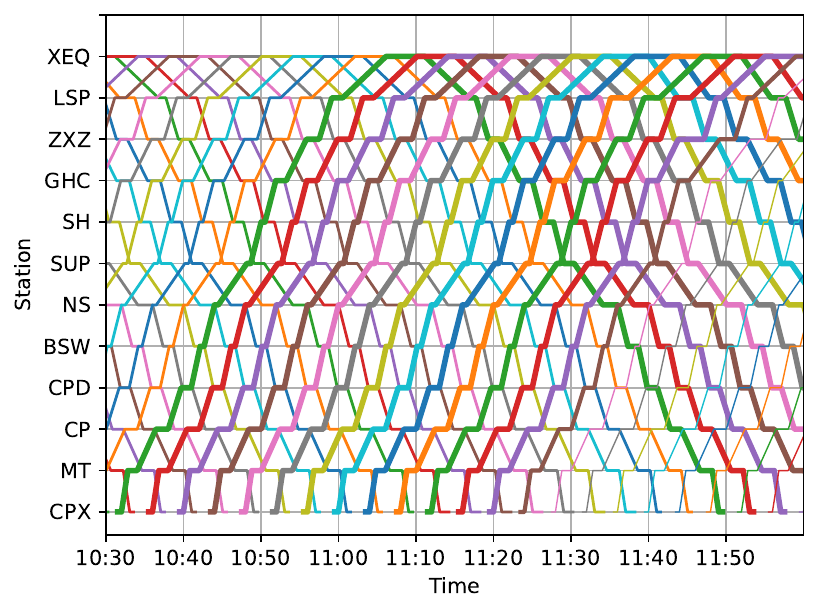}    
\caption{Timetable for Changping Line of the MILP approach} 
\label{lmpc-milp_timetable}
\end{center}
\end{figure}
\begin{figure}[htbp]
\begin{center}
\includegraphics[width=8.5cm]{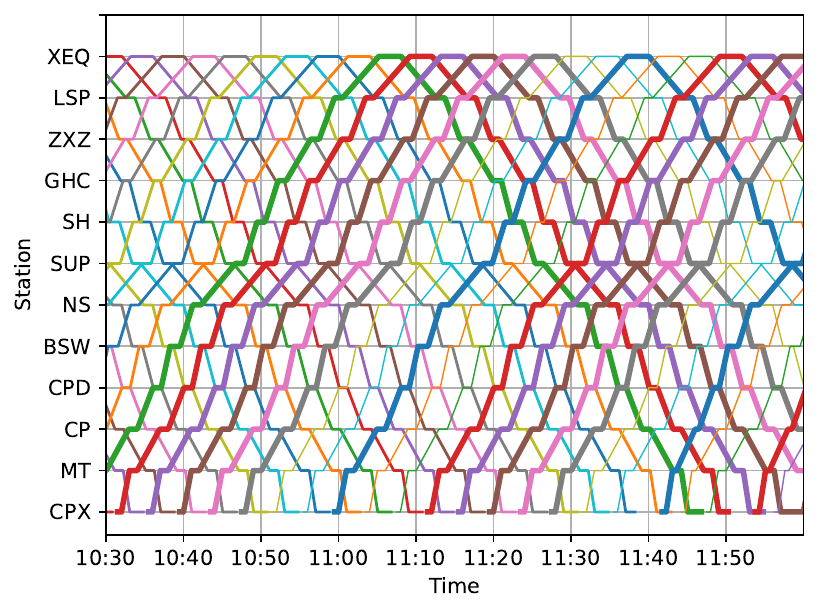}    
\caption{Timetable for Changping Line of the learning-based NLP approach} 
\label{lmpc-nlp_timetable}
\end{center}
\end{figure}
\begin{figure}[htbp]
\begin{center}
\includegraphics[width=8.5cm]{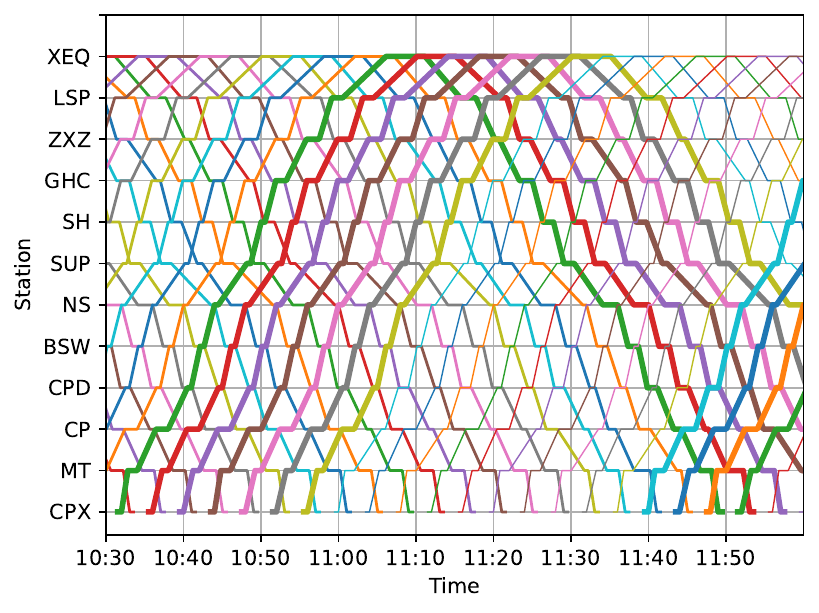}    
\caption{Timetable for Changping Line of the learning-based LP approach} 
\label{lmpc-lp_timetable}
\end{center}
\end{figure}

As an illustrative example, Fig.~\ref{lmpc-MPC_cost} presents the simulation results for one episode, showing the objective function values at each step for the Benchmark, warm-start MINLP, MILP, learning-based NLP, and learning-based LP approaches. The episode begins at 10:30 and spans 30 MPC steps. Fig.~\ref{lmpc-MPC_cost} indicates that the developed learning-based approaches achieve comparable performance concerning objective function values. The timetables for the Changping Line, covering from Station CPX to Station XEQ, are depicted in Fig.\ref{lmpc-benchmark_timetable}--Fig.~\ref{lmpc-lp_timetable} for the benchmark approach, the warm-start MINLP, MILP, learning-based NLP, and learning-based LP approaches, respectively, where the line width indicates the number of train units included in each train service.  
From Fig.\ref{lmpc-benchmark_timetable} to Fig.\ref{lmpc-lp_timetable}, it can be observed that all approaches deploy long trains from 10:30 to 10:50 to transport passengers. After 10:50, the benchmark, warm-start MINLP, and MILP approaches continue using long trains, whereas the learning-based NLP approach initially uses short trains before switching to long trains, resulting in a slight difference in the objective function value shown in Fig.\ref{lmpc-MPC_cost}. In contrast, the learning-based LP approach continues using short trains until 11:40, resulting in the worst performance in Fig.\ref{lmpc-MPC_cost}.

In general, the rail operator expects to generate a timetable as soon as possible to ensure efficient reasonable rolling stock circulation and crew scheduling, enabling effective management of the real-time traffic situations and passenger flows. The simulation results indicate that the developed learning-based MPC approach significantly reduces solution times with an acceptable trade-off in objective function performance. The developed learning-based MPC approach can generate timetables in real time with the train composition and rolling stock circulation plan generated by learning and the detailed departure and arrival times generated by nonlinear or linear programming. It should be noted that solving MINLP or MILP problems typically requires commercial solvers, such as \texttt{gurobi} and \texttt{cplex}, to generate solutions in a computationally efficient way. 
By generating integer variables with well-trained learning agents at each MPC step, the resulting continuous linear or nonlinear programming problems can be efficiently solved using a variety of available solvers and algorithms. 








\section{Conclusions}\label{lmpc-section7}
In this paper, we have investigated the passenger-oriented train rescheduling problem considering flexible train composition and rolling stock circulation. The passenger-oriented train rescheduling model of \cite{liu2023modeling} has been extended to include train compositions and rolling stock circulation considering time-varying passenger demands. 
To improve the online computational efficiency of model predictive control, we have combined the optimization-based and learning-based approaches, where the learning-based approach obtains integer variables, i.e., train compositions and train orders, by using pre-trained long short-term memory networks; then, the detailed timetables are optimized by solving a constrained optimization problem with the fixed integer variables. We have developed several presolve techniques to prune the subset of integer decision variables. Simulation results show that the developed learning-based framework can achieve comparable performance compared to the exact approach with an acceptable loss of feasibility while the solution time is significantly reduced. 

In the future, we will investigate the integration of reinforcement learning (instead of deep learning) and model predictive control for real-time train rescheduling to improve the learning ability of the approach. A state reduction approach can also be applied to improve performance and reduce memory usage. Among several directions, multi-agent learning-based approaches can also be a promising research direction for handling large-scale urban rail transit networks. 

\section*{Acknowledgements}
This research has received funding from the 
European Research Council (ERC) under the European Union’s Horizon 
2020 research and innovation programme (Grant agreement No. 101018826 - CLariNet). The work of X. Liu is also supported by the China Scholarship Council under Grant 202007090003.

\bibliography{mybibfile}

\end{document}